\documentclass[reprint,twocolumn,aip,superscriptaddress]{revtex4-1}
\usepackage[utf8]{inputenc}
\usepackage[T1]{fontenc}
\usepackage{amsmath}
\usepackage{amsfonts}
\usepackage{physics}
\usepackage[caption=false]{subfig}
\usepackage{ulem}
\usepackage[version=4]{mhchem}
\usepackage{bm}
\usepackage{amssymb}
\usepackage{mathtools}
\usepackage[mathscr]{eucal}
\usepackage{graphicx}
\usepackage[dvipsnames]{xcolor}
\usepackage[hidelinks,colorlinks=true,allcolors=Blue]{hyperref}

\begin{document}
\title{Spin Polarization from Circularly Polarized Light Induced Charge Transfer}
\author{Sindhana Pannir-Sivajothi}
\affiliation{Department of Chemistry, University of
California, Berkeley, California 94720, United States \looseness=-1}
\author{David T. Limmer}
\email{dlimmer@berkeley.edu}
\affiliation{Department of Chemistry, University of
California, Berkeley, California 94720, United States \looseness=-1}
\affiliation{Kavli Energy Nanoscience Institute at Berkeley, Berkeley, California 94720, United States \looseness=-1}
\affiliation{Chemical Sciences Division, Lawrence Berkeley National Laboratory, Berkeley, California 94720, United States \looseness=-1}
\affiliation{Material Sciences Division, Lawrence Berkeley National Laboratory, Berkeley, California 94720, United States \looseness=-1}
\begin{abstract}
We show how a spin polarization can be generated through the photo-induced electron transfer of an achiral donor-acceptor complex following chiral light excitation. In particular, we illustrate the basic energetic and symmetry requirements for chirality induced spin selectivity where the chirality emerges from the electronic degrees of freedom following excitation with circularly polarized light. We study this effect in a simple model of a  metalloporphyrin complex with an axial acceptor ligand using quantum mechanical rate theories and numerical simulations. We find that the spin polarization emerges due to the selective excitation of a ring current within the porphryin, breaking the degeneracy of the two degenerate spin states. The resultant spin polarization increases with the spin orbit coupling between the metal in the porphyrin and the axial ligand, and is transient, with a lifetime dependent on the rate of dephasing from the Jahn-Teller distortion mode. This proposed effect should be observable in spin-resolved photoemission spectroscopy.
\end{abstract}
\maketitle

The chirality-induced spin selectivity (CISS) effect, where electron transmission through chiral molecules is spin-dependent, has been observed in both photoelectron transport through ground-state chiral molecules \cite{ray1999asymmetric,gohler2011spin} and photoinduced electron transfer.\cite{eckvahl2023direct} 
 CISS is typically attributed to chirality defined by nuclear geometry, however achiral donor-acceptor systems excited with circularly polarized light (CPL) can admit transient chirality when both nuclear and  electronic degrees of freedom are considered.\cite{chen2024chiral,moitra2025light} This suggests that a transient spin polarization may be generated following CPL in achiral complexes. In this Letter, we illustrate that a spin polarization in CPL-driven electron transfer in achiral donor-acceptor complexes is indeed possible transiently and discuss the underlying mechanism. This presents a novel means of controlling electron spin in molecular systems with potential application for molecular qubits.\cite{ferrando2016modular,wasielewski2020exploiting,santanni2024metalloporphyrins}

\begin{figure}
	\includegraphics[width=\columnwidth]{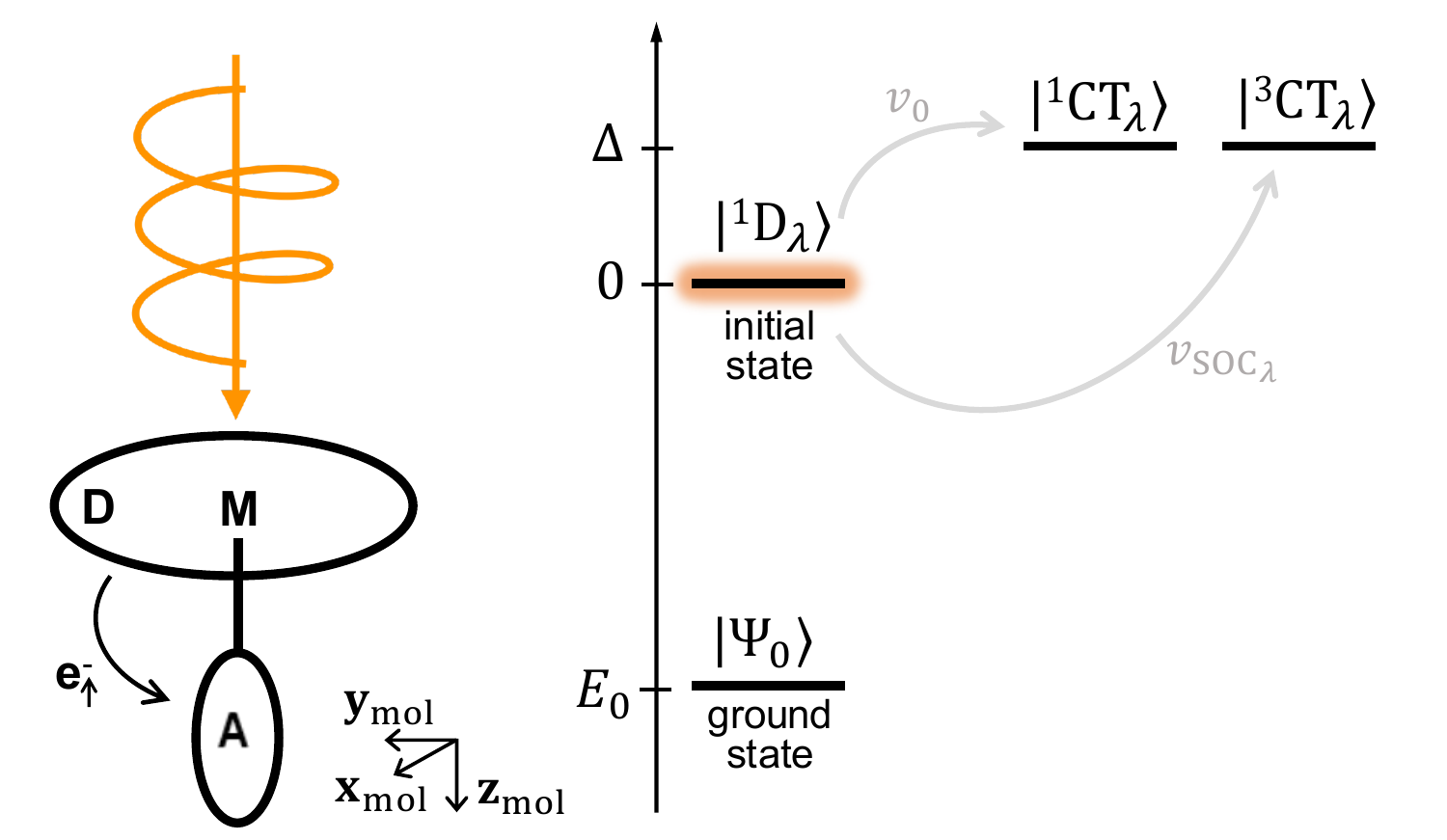}
	\caption{\label{fig:schem} Electron transfer in an achiral donor-acceptor complex driven by circularly polarized light. Left- and right-circularly polarized light excite the donor singlet states $\ket{^1\mathrm{D}_{\lambda}}\equiv$ $^1$[D$^*_{\lambda}$-A] with $\lambda=\pm$ from the ground state $\ket{\Psi_0}\equiv$ $^1$[D-A]. Singlet and triplet charge transfer states $\ket{^1\mathrm{CT}_{\lambda}}\equiv$ $^1$[D$^{\bullet+}_{\lambda}$-A$^{\bullet-}$] and $\ket{^3\mathrm{CT}_{\lambda}}\equiv$ $^3$[D$^{\bullet+}_{\lambda}$-A$^{\bullet-}$] are nearly degenerate and detuned from $\ket{^1\mathrm{D}_{\lambda}}$ by $\Delta$, while $v_0$ and $v_{\mathrm{SOC}_{\lambda}}$ mediate spin-conserving and spin-orbit coupled electron transfer.}
\end{figure}

When a molecule absorbs CPL, the photon's angular momentum can be transferred to its electronic orbital degrees of freedom, producing excited states with nonzero projection of orbital angular momentum along the light axis. This requires electric-dipole-allowed excited states capable of carrying orbital angular momentum. In molecules, such states arise from degenerate manifolds belonging to multidimensional irreducible representations. Examples include $\pi$-conjugated molecules such as metalloporphyrins\cite{barth2006unidirectional} and metallophthalocyanines,\cite{Isago2015,claessens2008phthalocyanines} as well as monolayer transition metal dichalcogenides.\cite{cao2012valley}

For concreteness, we consider an achiral donor molecule with CPL-addressable degenerate excited states $\ket{^1\mathrm{D}_{\lambda}}$ where the label $\lambda=\pm$ depends on its orbital angular momentum, as sketched in Fig. \ref{fig:schem}. We define a molecular frame ($\mathbf{x}_{\mathrm{mol}},\mathbf{y}_{\mathrm{mol}},\mathbf{z}_{\mathrm{mol}}$), with $\mathbf{z}_{\mathrm{mol}}$ pointing from donor to acceptor and the molecule is oriented such that the light axis aligns with $\mathbf{z}_{\mathrm{mol}}$ as in Fig.~\ref{fig:schem}. To understand the subsequent charge-transfer dynamics, we construct a minimal electronic Hamiltonian for this donor--acceptor system using only its frontier orbitals. Within this framework, two charge-transfer pathways emerge: a spin-conserving channel ($\ket{^1\mathrm{D}_{\lambda}}\to\ket{^1\mathrm{CT}_{\lambda}}$), and a spin--orbit-coupled channel ($\ket{^1\mathrm{D}_{\lambda}}\to\ket{^3\mathrm{CT}_{\lambda}}$) that depends on the helicity of the exciting CPL. The model predicts a net spin polarization that reverses under $\lambda \to -\lambda$ and grows with the spin--orbit coupling (SOC) strength $v_{\mathrm{SOC}_{\lambda}}$. This spin polarization decays due to vibronic coupling in this case originating in Jahn-Teller distortions that lift the $\lambda=\pm$ degeneracy,\cite{hougen1964vibronic,panariti2025long} which we model using Bloch-Redfield equations. The transient generation of a spin polarization therefore, occurs when the charge-transfer rate exceeds the rate of $\lambda=\pm$ orbital angular momentum state mixing. Since such ultrafast coherent charge transfer has been observed in dyads with a metalloporphyrin donor and an axially bound acceptor,\cite{benitz2021photoinduced,guragain2023zinc,chen2024ultra} we take this as a representative system for our study but expect the basic mechanism to hold generally.

To build a minimal Hamiltonian, we consider complex-valued frontier orbitals for the donor porphyrin, with creation operators $\hat{d}_{\mathrm{H}_{\pm}\sigma}^{\dagger}$ for the HOMOs and $\hat{d}_{\mathrm{L}_{\pm}\sigma}^{\dagger}$ for the LUMOs. These molecular orbitals carry a finite projection of orbital angular momentum along $\mathbf{z}_{\mathrm{mol}}$. For the acceptor, we have real-valued LUMOs with creation operators $\hat{a}_{\mathrm{L}\sigma}^{\dagger}$, where $\sigma\in \{\alpha,\beta\}$ denotes spin. The complex donor orbitals are constructed from Gouterman's four-orbital model, the canonical model for porphyrin optical spectra.\cite{gouterman1961spectra} The Gouterman model includes two exactly degenerate LUMOs and two accidentally degenerate HOMOs all of which are real-valued and have transition dipole moments along $\mathbf{x}_{\mathrm{mol}}$ and $\mathbf{y}_{\mathrm{mol}}$. We construct complex frontier orbitals, $\hat{d}_{\mathrm{H}_{\pm}\sigma}^{\dagger}$ and $\hat{d}_{\mathrm{L}_{\pm}\sigma}^{\dagger}$, as linear combinations of the degenerate Gouterman orbitals and singly excited states built from these yield transition dipoles $\mathbf{x}_{\mathrm{mol}}\pm i\mathbf{y}_{\mathrm{mol}}$, which are selectively addressable with CPL (see Supplementary S1).
When donor-acceptor dyads have orthogonal $\pi$-systems, they can exhibit substantial SOC-assisted charge transfer,\cite{dance2008intersystem,fay2024unraveling} therefore, we choose to study a $\pi$-conjugated dyad with perpendicular donor  and acceptor $\pi$-planes. Since the acceptor is bound to the donor axially via a metal center, the metal forms part of both donor and acceptor $\pi$-systems and charge transfer occurs via it.

Using the frontier orbitals and the electronic ground state $\ket{\Psi_0}$ of the donor-acceptor dyad, we define the donor singlet excited states as 
\begin{equation}\label{eq:D1}
    \ket{^1\mathrm{D}_{\lambda}}=\frac{1}{\sqrt{2}}(\hat{d}_{\mathrm{L}_{\lambda}\alpha}^{\dagger}\hat{d}_{\mathrm{H}_{\lambda}\alpha}+\hat{d}_{\mathrm{L}_{\lambda}\beta}^{\dagger}\hat{d}_{\mathrm{H}_{\lambda}\beta})\ket{\Psi_0},
\end{equation}
 where $\lambda = \pm$ corresponds to states excited with left- and right-CPL (see Supplementary S1). These will be our initial states. The donor excited states, $\ket{^1\mathrm{D}_{+}}$ and $\ket{^1\mathrm{D}_{-}}$, constitute an enantiomeric pair linked by reflection about a mirror plane.
Although the donor lacks continuous rotational symmetry about $\mathbf{z}_{\mathrm{mol}}$, its discrete symmetry (\textit{e.g.}, $D_{4h}$ in metalloporphyrins) supports quasi-angular-momentum states that are approximate eigenstates of the z-component of orbital angular momentum $\hat{L}_{z_{\mathrm{mol}}}=\mathbf{\hat{L}}\cdot \mathbf{z}_{\mathrm{mol}}$ with $\hat{L}_{z_{\mathrm{mol}}}\ket{^1\mathrm{D}_{\lambda}}\approx \lambda m\hbar \ket{^1\mathrm{D}_{\lambda}}$ (see Supplementary S2). These quasi-angular momentum states are selectively excitable with CPL and carry opposite ring currents for $\lambda=\pm$.\cite{malley1968zeeman,barth2006unidirectional} The donor-acceptor complex possesses $C_{4v}$ symmetry, lower than the $D_{4h}$ symmetry of the donor alone, which includes vertical mirror planes $\sigma_v$ containing the principal molecular axis $\mathbf{z}_{\mathrm{mol}}$. Under reflection about such a plane, the orbital angular momentum transforms as $\sigma_v\hat{L}_{z_{\mathrm{mol}}}\sigma_v^{-1}=-\hat{L}_{z_{\mathrm{mol}}}$ as it is the component of an axial vector operator parallel to the mirror plane. Consequently, the excited states are related to each other by reflection about a mirror plane $\sigma_v\ket{^1\mathrm{D}_{+}}\propto\ket{^1\mathrm{D}_{-}} $,\cite{neufeld2019background} analogous to enantiomers of chiral molecules, and are non-superimposable by proper rotations in three dimensions. We will refer to these chiral states as \textit{transient enantiomers}. These transient enantiomers do not represent ‘true chirality,’ despite being non-superimposable mirror images, but instead belong to the class of false chiral states because they are also related by time-reversal symmetry.\cite{barron1986true,ordonez2019propensity} Consequently, observables that are even under time reversal, such as the charge density, are identical for the two enantiomers, whereas observables that are odd under time reversal, such as spin polarization, acquire opposite values.

 In a randomly oriented ensemble of achiral donor–acceptor complexes, the state prepared by CPL depends on the orientation of the molecular axis $\mathbf{z}_{\mathrm{mol}}$ relative to the photon angular momentum direction $\mathbf{z}_{\mathrm{lab}}$. For a given orientation, CPL prepares the molecule in a linear combination of $\ket{^1\mathrm{D}_{+}}$ and $\ket{^1\mathrm{D}_{-}}$, where molecules with orientation $\mathbf{z}_{\mathrm{mol}}\cdot\mathbf{z}_{\mathrm{lab}}>0$ have a larger $\ket{^1\mathrm{D}_{+}}$ component and those with $\mathbf{z}_{\mathrm{mol}}\cdot\mathbf{z}_{\mathrm{lab}}<0$ have a larger $\ket{^1\mathrm{D}_{-}}$ component. Upon ensemble averaging, the populations of the two enantiomers are equal, resulting in a racemic mixture. However, because CPL transfers angular momentum to the molecule and thereby defines a preferred direction in the laboratory frame, the ensemble is analogous to an oriented racemic mixture rather than an isotropic one. Consequently, the spin polarization, when projected onto each molecule’s axis and then averaged over the ensemble, vanishes, whereas its ensemble-averaged projection onto the laboratory axis $\mathbf{z}_{\mathrm{lab}}$ remains non-zero.

The singlet and $M_S=0$ triplet charge-transfer (CT) states are 
\begin{subequations}\label{eq:CT}
\begin{align}
     \ket{^1\mathrm{CT}_{\lambda}}=\frac{1}{\sqrt{2}}(\hat{a}_{\mathrm{L}\alpha}^{\dagger}\hat{d}_{\mathrm{H}_{\lambda}\alpha}+\hat{a}_{\mathrm{L}\beta}^{\dagger}\hat{d}_{\mathrm{H}_{\lambda}\beta})\ket{\Psi_0},\\
    \ket{^3\mathrm{CT}_{\lambda}}=\frac{1}{\sqrt{2}}(\hat{a}_{\mathrm{L}\alpha}^{\dagger}\hat{d}_{\mathrm{H}_{\lambda}\alpha}-\hat{a}_{\mathrm{L}\beta}^{\dagger}\hat{d}_{\mathrm{H}_{\lambda}\beta})\ket{\Psi_0}.
\end{align}
\end{subequations}
The CT state retains the label $\lambda=\pm$ because the donor hole carries a ring current, resulting in finite expectation value of  $\hat{L}_{z_{\mathrm{mol}}}$. The singlet CT state is detuned by $\Delta$ from the singlet donor excited state. The singlet charge transfer state $\ket{^1\mathrm{CT}_{\lambda}}$ is coupled to the donor singlet state $\ket{^1\mathrm{D}_{\lambda}}$ through a diabatic coupling with magnitude ${v}_0$ mediate by the metal center.  Due to the orientation of the $\pi$-planes, for the donor we include only the $p_z$ atomic orbitals of its atoms and for the acceptor, only the $p_x$ orbitals.  We include the metal's $d_{xz}$ and $d_{yz}$ orbitals, where $d_{xz}$ contributes to both $\pi$-systems and thus can mediate charge transfer directly. 
\color{black}

To account for the SOC-assisted charge transfer, we include the mean-field Breit-Pauli form of the spin-orbit coupling.\cite{marian2012spin} Within our minimal model basis, only the metal center contributes to the one-center SOC.\cite{gouterman1973porphyrins,ake1969porphyrins} The one-center SOC terms are defined as matrix elements where the angular momentum operator and both atomic orbitals are centered around the same nucleus. Because the angular momentum operators are pure imaginary operators and have zero diagonal matrix elements in a real-valued atomic orbital basis and since we retain only one atomic orbital per atom except for the metal, it is evident that only the metal center will contribute to the one-center SOC. Therefore, the diabatic states are additionally coupled through a single-electron SOC term of the Hamiltonian with magnitude ${v}_{\mathrm{SOC}_{\lambda}}$,  as derived in Supplementary S2 using the mean-field SOC and H\"{u}ckel theory. 

\color{black}
The resultant electronic Hamiltonian, $\hat{H}_\mathrm{e}$, spanned by $\ket{^1\mathrm{D}_{\lambda}}$, $\ket{^1\mathrm{CT}_{\lambda}}$, and $\ket{^3\mathrm{CT}_{\lambda}}$ is thus
\begin{eqnarray}\label{eq:He}
\begin{aligned}  \hat{H}_{\mathrm{e}}=\sum_{\lambda=+,-} & \Delta \left ( \ket{^1\mathrm{CT}_{\lambda}}\bra{^1\mathrm{CT}_{\lambda}}+\ket{^3\mathrm{CT}_{\lambda}}\bra{^3\mathrm{CT}_{\lambda}} \right )\\
&+v_0 \left ( \ket{^1\mathrm{D}_{\lambda}}\bra{^1\mathrm{CT}_{\lambda}}+\ket{^1\mathrm{CT}_{\lambda}}\bra{^1\mathrm{D}_{\lambda}} \right ) \\
&+v_{\mathrm{SOC}_{\lambda}} \left ( \ket{^1\mathrm{D}_{\lambda}}\bra{^3\mathrm{CT}_{\lambda}}+\ket{^3\mathrm{CT}_{\lambda}}\bra{^1\mathrm{D}_{\lambda}} \right )
\end{aligned}
\end{eqnarray}
where all energy differences are with respect to the donor singlet. Since this electronic Hamiltonian does not couple states with different $\lambda$, an initial state prepared in a given angular momentum subspace $\lambda$ remains there under time evolution with $\hat{H}_{\mathrm{e}}$. Note $v_{\mathrm{SOC}_{\lambda}}$ depends on helicity with $v_{\mathrm{SOC}_{-}}=-v_{\mathrm{SOC}_{+}}$. One can interpret this difference in sign of the SOC matrix element as due to an effective transient magnetic field produced by the ring current upon excitation with CPL. Previous work suggests that such ultrafast currents may generate experimentally observable magnetic fields.\cite{yuan2018generation,de2023tunable} The spin-conserving electron transfer occurs solely via the $d_{xz}$ orbital while the SOC charge transfer involves both the $d_{xz}$ and $d_{yz}$ orbitals of the metal center. Unlike previous work where the spin-preserving electron transfer matrix element is real and the SOC matrix element is purely imaginary,\cite{fay2021origin}  in our case both the direct and SOC-assisted electron transfer matrix elements are real-valued as we have complex frontier orbitals with finite angular momentum. 

The triplet donor states with spin projection $M_S=0$ along the quantization axis $\mathbf{z}_{\mathrm{mol}}$ are
  \begin{equation}\label{eq:D3}
      \ket{^3\mathrm{D}_{\lambda}}=\frac{1}{\sqrt{2}}(\hat{d}_{\mathrm{L}_{\lambda}\alpha}^{\dagger}\hat{d}_{\mathrm{H}_{\lambda}\alpha}-\hat{d}_{\mathrm{L}_{\lambda}\beta}^{\dagger}\hat{d}_{\mathrm{H}_{\lambda}\beta})\ket{\Psi_0}.
  \end{equation}
In a mean-field Hamiltonian, the singlet and triplet states are degenerate. However, the full Hamiltonian would include additional exchange interaction terms of the form $J\sum_{\sigma=\alpha,\beta}\sum_{\tau=\alpha,\beta}\hat{d}^{\dagger}_{\mathrm{L}_{\lambda}\sigma}\hat{d}^{\dagger}_{\mathrm{H}_{\lambda}\tau}\hat{d}_{\mathrm{L}_{\lambda}\tau}\hat{d}_{\mathrm{H}_{\lambda}\sigma}$ that lower the energy of the triplet with respect to the singlet. For donor excited states, the singlet-triplet energy gap $2J \sim 400\,\mathrm{meV}$ for metalloporphyrins.\cite{kee2008photophysical}
In the CT configuration, the electron and hole are spatially separated, so the exchange interaction is small and the singlet and triplet CT states are nearly degenerate.\cite{weiss2003direct} For our calculations we use $\hat{H}_{\mathrm{e}}$ as defined in Eq. \ref{eq:He}, where we take the energy difference between the singlet and triplet CT states to be zero as it will be much smaller than other energy scales in the problem, and we exclude the state $\ket{^3\mathrm{D}_{\lambda}}$ from our analysis as it will be far-detuned  from $\ket{^1\mathrm{D}_{\lambda}}$, $\ket{^1\mathrm{CT}_{\lambda}}$, and $\ket{^3\mathrm{CT}_{\lambda}}$. The electronic Hamiltonian with $\ket{^3\mathrm{D}_{\lambda}}$ included is in Supplementary S2.

While the donor singlets and $M_S=0$ triplets $\ket{^{1(3)}\mathrm{D}_{\lambda}}$ are eigenstates of both $\hat{S}_{\mathrm{D}z_{\mathrm{mol}}}$ and $\hat{S}_{\mathrm{A}z_{\mathrm{mol}}}$ with eigenvalue $0$, the charge transfer states are not.  The total $\mathbf{z}_{\mathrm{mol}}$ projected spin $\hat{S}_{z_{\mathrm{mol}}}=\hat{S}_{\mathrm{Dz_{\mathrm{mol}}}}+\hat{S}_{\mathrm{Az_{\mathrm{mol}}}}$, where the spin of electrons localized on the donor is
     \begin{equation}
     \begin{aligned}
         \hat{S}_{\mathrm{D}z_{\mathrm{mol}}}=&\hbar/2\sum_{\lambda=+,-}(\hat{d}_{\mathrm{H}_{\lambda}\alpha}^{\dagger}\hat{d}_{\mathrm{H}_{\lambda}\alpha}-\hat{d}_{\mathrm{H}_{\lambda}\beta}^{\dagger}\hat{d}_{\mathrm{H}_{\lambda}\beta} \\
     &+\hat{d}_{\mathrm{L}_{\lambda}\alpha}^{\dagger}\hat{d}_{\mathrm{L}_{\lambda}\alpha}-\hat{d}_{\mathrm{L}_{\lambda}\beta}^{\dagger}\hat{d}_{\mathrm{L}_{\lambda}\beta})
     \end{aligned}
     \end{equation}
 and of those in the acceptor is 
 \begin{equation}
\hat{S}_{\mathrm{A}z_{\mathrm{mol}}}=\hbar/2(\hat{a}_{\mathrm{L}\alpha}^{\dagger}\hat{a}_{\mathrm{L}\alpha}-\hat{a}_{\mathrm{L}\beta}^{\dagger}\hat{a}_{\mathrm{L}\beta}).    
 \end{equation} 
 The spin polarization is defined as $ \hat{P}_{z}=\hat{S}_{\mathrm{A}z_{\mathrm{mol}}}-\hat{S}_{\mathrm{D}z_{\mathrm{mol}}}$ and will be the primary observable of our interest.
In the representation of our electronic states, the spin polarization manifests as the real part of the coherence between the charge transfer singlets and triplets,
\begin{equation}
\hat{P}_{z}=\sum_{\lambda=+,-}\Big(\hbar\ket{^3\mathrm{CT}_{\lambda}}\bra{^1\mathrm{CT}_{\lambda}}+\mathrm{h.c.}\Big)
\end{equation}
as has been noted in previous work.\cite{fay2021origin}

We want to model ultrafast coherent electron transfer and will not explicitly include a vibrational mode that is displaced between the donor excited state and the charge transfer state, so our analysis works best for rigid donor-acceptor complexes with minimal outer and inner-sphere reorganization, small detunings $\Delta$, and large electronic couplings that facilitate this ultrafast charge transfer process.\cite{ortiz2017through,pieslinger2022excited,kang2019orientational} 
Additionally, the electron transfer timescale, set by the coupling strength $\sqrt{v_0^2+v_{\mathrm{SOC}_\pm}^2}$, must be shorter than the vibronic mixing timescale between the $\lambda=\pm$ states for this effect to be experimentally observable, as quantified later in the manuscript. For our calculations, we use $v_0=75\,$meV which is reasonable for systems that satisfy the above criteria,\cite{ortiz2017through} 
\color{black} and we use $v_{\mathrm{SOC}_{\pm}}=\pm25\,$meV  based on SOC matrix elements in donor-acceptor complexes with heavy transition metal atoms.\cite{li2005synthetic,gomez2025computational} 

\begin{figure}
	\includegraphics[width=\linewidth]{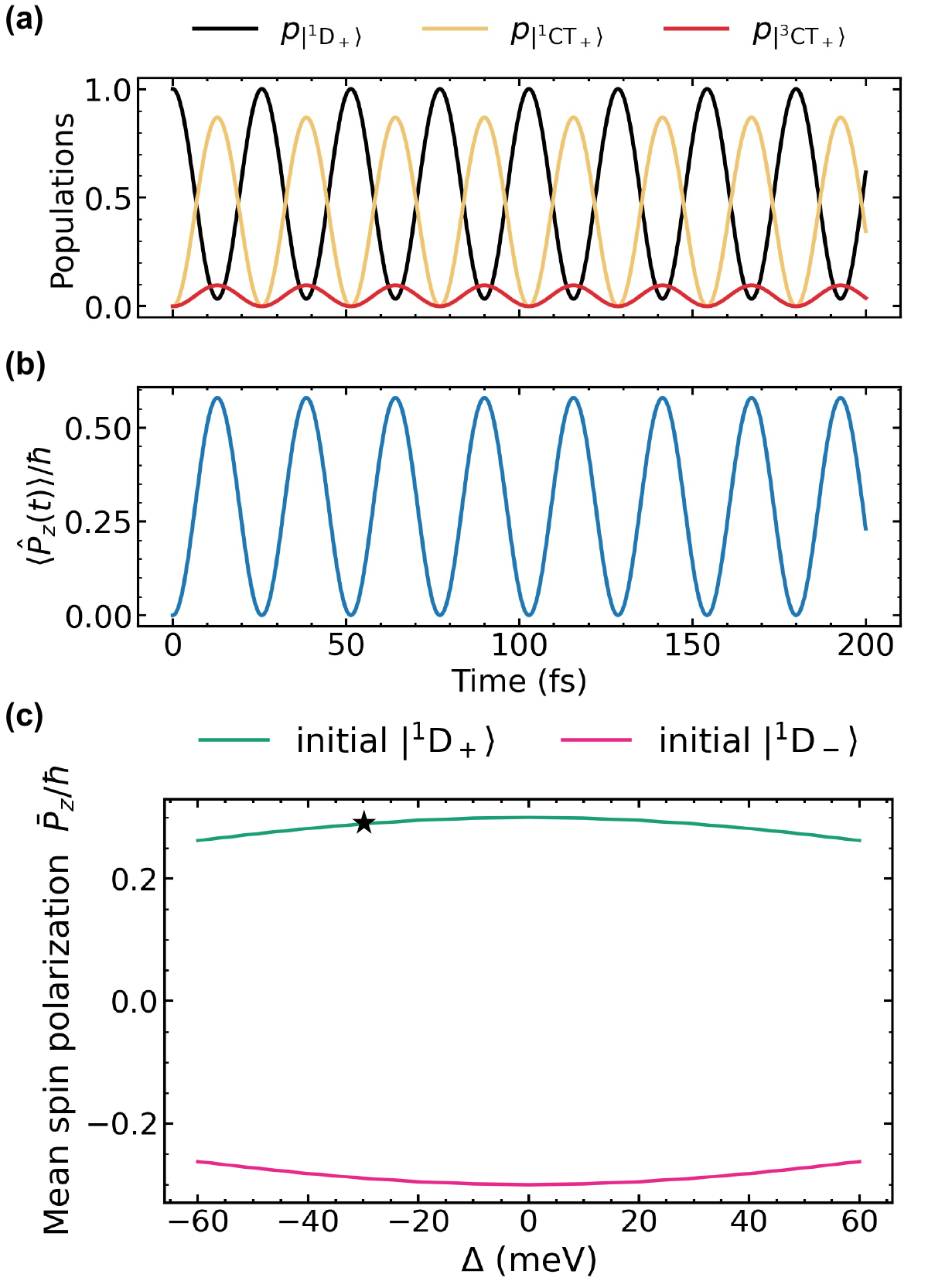}
	\caption{\label{fig:fig2} Unitary dynamics. (a) The populations in the singlet donor $p_{\ket{^1\mathrm{D}_+}}$ (black), singlet charge transfer $p_{\ket{^1\mathrm{CT}_+}}$ (yellow) and the triplet charge transfer $p_{\ket{^3\mathrm{CT}_+}}$ (red) states and (b) the spin polarization $\langle\hat{P}_{z}(t)\rangle$ following excitation into $\ket{^1\mathrm{D}_{+}}$. (c) Mean spin polarization $\bar{P}_{z}$ for different detunings $\Delta$ and its dependence on initially excited state. Black star  indicates condition in (a) and (b).}
\end{figure}

Given an initial state $\ket{^1\mathrm{D}_{+}}$ the subsequent unitary dynamics for the population and spin polarization can be deduced exactly. These are shown in Fig. \ref{fig:fig2}(a-b). The population of the donor, $ p_{\ket{^1\mathrm{D}_+}}(t)=|\bra{^1\mathrm{D}_{+}(t)} \ket{^1\mathrm{D}_{+}(0)}|^2$, singlet charge transfer state, $p_{\ket{^1\mathrm{CT}_+}}(t)=|\bra{^1\mathrm{CT}_+(t)} \ket{^1\mathrm{D}_{+}(0)}|^2$ and triple charge transfer state, $p_{\ket{^3\mathrm{CT}_+}}(t)=|\bra{^3\mathrm{CT}_+(t)} \ket{^1\mathrm{D}_{+}(0)}|^2$ each beat with a Rabi frequency, $\Omega=\sqrt{(\Delta/2)^2+v_0^2+v_{\mathrm{SOC}_{+}}^2}$. The donor state is exactly out of phase with the two charge transfer states, whose relative amplitudes depend on the ratio of $v_0^2/v^2_{\mathrm{SOC}_+}$.  The spin polarization has a time dependence that is given by
\begin{equation}
\langle \hat{P}_{z}(t) \rangle=\frac{\hbar v_0v_{\mathrm{SOC}_{+}}}{\Omega^2}\Big[1-\cos(2\Omega t/\hbar)\Big],
\end{equation}
which we plot in Fig.~\ref{fig:fig2}(b). The spin polarization beats with the same Rabi frequency as the populations, and with an amplitude proportional to the size of the spin-orbit coupling. The coupling parameters $v_0$ and $v_{\mathrm{SOC}_{\pm}}$ depend on the metal $d_{\pi}$ character of the donor and acceptor LUMOs. Consequently, geometric factors such as donor–acceptor separation influence the spin polarization by modifying this $d_{\pi}$ character and thus these couplings.
The mean spin polarization $ \bar{{P}_z}=\lim_{\tau\to\infty}\frac{1}{\tau}\int_{0}^{\tau}dt \langle \hat{P}_{z}(t)\rangle$ is non-zero and takes on opposite sign depending on whether the initial state is $\ket{^1\mathrm{D}_+}$ or $\ket{^1\mathrm{D}_-}$
\begin{equation}\label{eq:meanDeltaPz}
\begin{aligned}
     \bar{{P}_z}=&\hbar\frac{v_0v_{\mathrm{SOC}_{\lambda}}}{\Omega^2},
\end{aligned}
\end{equation}
 that is, the mean spin-polarization flips sign when excited with right- versus left-circularly polarized light as $v_{\mathrm{SOC}_{+}}=-v_{\mathrm{SOC}_{-}}$. We plot $\bar{P}_z$ in Fig. \ref{fig:fig2}(c), when the initial state is $\ket{^1\mathrm{D}_+}$ and $\ket{^1\mathrm{D}_-}$ as a function of $\Delta$, the energy difference between the donor and charge transfer states. We find $\sim0.3\hbar$ mean spin polarization at $\Delta=0$.

Ideal systems to test our predictions in Fig.~\ref{fig:fig2}(c) are donor-acceptor dyads with identical donor and acceptor molecules ($\Delta =0$). By introducing substituents on the acceptor, its LUMO--and hence $\Delta$--can be tuned.\cite{ma2009substituent,barbee2012revealing} Across such a series, a CPL pump followed by spin-resolved photoemission spectroscopy would provide an experimental plot of mean spin-polarization like Fig.~\ref{fig:fig2}(c). However, in reality the initial excitation will dephase through vibronic coupling. Specifically, it is known that in metalloporphyrins and related systems, the degenerate $\ket{^1\mathrm{D}_+}$ and $\ket{^1\mathrm{D}_-}$ are coupled through Jahn-Teller distortions with the typical depolarization timescales of $\sim50-200\,\mathrm{fs}$.\cite{galli1993direct,hildner2011femtosecond,sun2022coherent} Since the spin polarization depends on the nature of the excited donor state, this dephasing should cause the mean spin polarization to decay. 

We quantify these spin relaxation rates using Bloch-Redfield equations, assuming that the electron-phonon coupling is small compared to $v_0$ and $v_{\mathrm{SOC}_\lambda}$.\cite{limmer2024statistical,jeske2015bloch,campaioli2024quantum} Typically, for degenerate electronic states $\ket{^1\mathrm{D}_{\pm}}$ which are quasi-eigenstates of the orbital angular momentum operator, the $(E\cross e)$ Jahn-Teller Hamiltonian is used where a pair of degenerate vibrational modes couple them.\cite{nandipati2021dynamical} However, for molecules like metalloporphyrin that have an approximate four-fold symmetry, the Jahn-Teller active vibrational modes are non-degenerate.\cite{hougen1964vibronic,panariti2025long} We model this by coupling the degenerate $\lambda=\pm$ orbital angular momentum states to a bath through a bilinear system-bath coupling of the form $\hat{V}_{\mathrm{e-b}}=\hat{V}_{\mathrm{e}}\otimes\hat{B}$ where
\begin{equation}
	\begin{aligned}
		\hat{V}_{\mathrm{e}}=& f_{\mathrm{D}}\ket{^1\mathrm{D}_+}\bra{^1\mathrm{D}_-}+ f_{\mathrm{CT}}\ket{^1\mathrm{CT}_+}\bra{^1\mathrm{CT}_-}\\
        &+f_{\mathrm{CT}}\ket{^3\mathrm{CT}_+}\bra{^3\mathrm{CT}_-}+\mathrm{h.c.},
	\end{aligned}
\end{equation}
with dimensionless coupling constants $f_{\mathrm{D}}$ and $f_{\mathrm{CT}}$. The bath operator $\hat{B}=\sum_k (g_k\hat{b}_k^{\dagger}+g_k^*\hat{b}_k)$ where $\hat{b}_k$ is the annihilation operator of the bath mode $k$ and the bath has a Drude-Lorentz \color{black} spectral density 
\begin{equation}
J(\omega)=\sum_k \frac{|g_k|^2}{\hbar} \delta(\omega-\omega_k)=\frac{2\Lambda\omega\omega_c}{\omega^2+\omega_c^2}
\end{equation}
with reorganization energy $\Lambda=0.2\,\mathrm{meV}$ so the coupling strength is much smaller than other energy scales in $\hat{H}_{\mathrm{e}}$ for a range of $f_{\mathrm{D}},f_{\mathrm{CT}}\in[-0.4,0.4]$. We consider a fast bath with cutoff frequency $\hbar\omega_c=30\,\mathrm{meV}$ ($1/\omega_c\approx22\,$fs), corresponding to a bath-correlation decay time in the typical condensed-phase range of $\sim20-150\,$fs. \cite{ishizaki2009theoretical,hu2025trajectory}

  \begin{figure}
	\includegraphics[width=\linewidth]{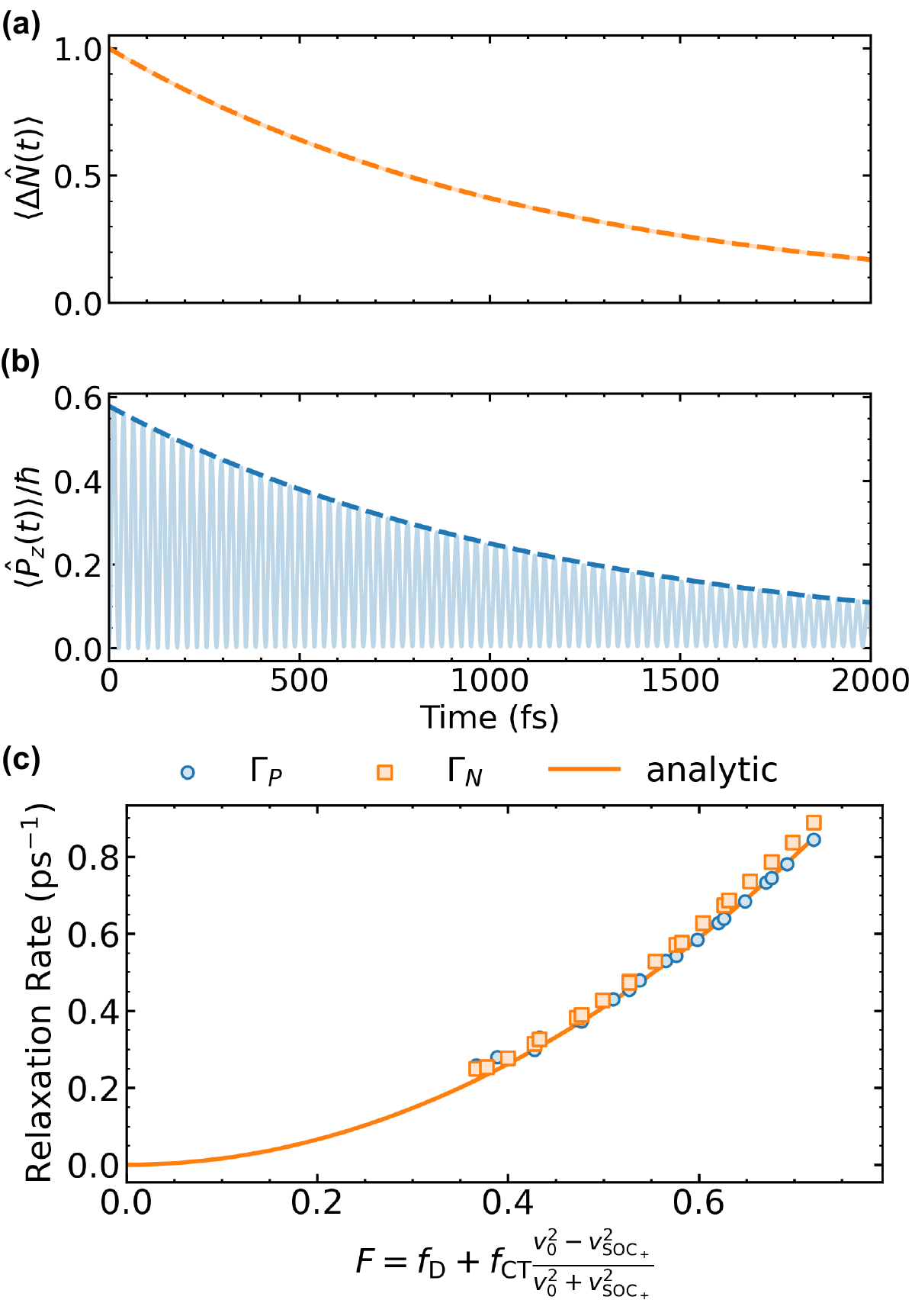}
	\caption{\label{fig:fig3} Relaxational dynamics. (a) Population difference $\langle \Delta \hat{N}(t)\rangle=\langle\hat{N}_+(t)\rangle-\langle\hat{N}_-(t)\rangle$ (solid light orange) and its exponential fit $\exp(-\Gamma_{N}t)$ (dashed orange) and (b) the spin polarization $\langle\hat{P}_{z}(t)\rangle$ (solid light blue) and the exponential fit to its envelope $\exp(-\Gamma_{P}t)$ (dashed blue)  for $\Delta=-30\,\mathrm{meV}$ and $f_{\mathrm{D}}=f_{\mathrm{CT}}=0.4$. (c)
    The decay rates $\Gamma_{P}$ (blue circles) and $\Gamma_{N}$ (orange squares) as a function of dimensionless coupling strengths $f_{\mathrm{D}}$ and $f_{\mathrm{CT}}$. The solid line is the expected rate from our analytic expression in Eq. \ref{Eq:popdif}.
    }
\end{figure}

Along with the spin polarization $\hat{P}_{z}$, we also study the relaxation rate of the difference in population between the two transient-enantiomers $\Delta \hat{N}=\hat{N}_+-\hat{N}_-$ where 
\begin{equation}
\hat{N}_{\lambda}=\ket{^1\mathrm{D}_{\lambda}}\bra{^1\mathrm{D}_{\lambda}}+\ket{^1\mathrm{CT}_{\lambda}}\bra{^1\mathrm{CT}_{\lambda}}+\ket{^3\mathrm{CT}_{\lambda}}\bra{^3\mathrm{CT}_{\lambda}}
\end{equation}
is the  population of the $\lambda$ enantiomer.
Within Redfield theory when the initial state is $\ket{^1\mathrm{D}_{+}}$, the approximate equation of motion of the population difference $\langle\Delta\hat{N}(t)\rangle$ has a simple form
\begin{equation}\label{Eq:popdif}
    \frac{d\langle\Delta \hat{N}(t)\rangle}{dt} \approx -\frac{2\pi k_\mathrm{B} T}{\hbar}\frac{\Lambda}{\hbar\omega_c}F ^2\langle\Delta\hat{N}(t)\rangle,
\end{equation}
  where $k_\mathrm{B} T$ is Boltzmann's constant times temperature. 
We derive this approximate form for the relaxation rates in Supplementary S4.  
  The decay rate is proportional to the reorganization energy and the temperature, and inversely proportional to the cutoff frequency. The rate also scales quadratically with an effective coupling strength, $F$, where 
  \begin{equation}
F =f_{\mathrm{D}}+f_{\mathrm{CT}}\frac{v_0^2-v_{\mathrm{SOC}_+}^2}{v_0^2+v_{\mathrm{SOC}_+}^2}
\end{equation}
is the effective system-bath coupling strength. 
  We compute the relaxation rates $\Gamma_{N}$ and $\Gamma_{P}$ of the population difference and spin polarization, respectively, numerically by fitting an exponential to the time series of $\langle\Delta\hat{N} (t)\rangle$ and to the envelope of $\langle \hat{P}_{z} (t)\rangle$ propagated with a Redfield equation of motion, and shown in Figs~\ref{fig:fig3}(a-b). Over a range of couplings, we find that both decay with a similar rate that is approximately given by Eq.~\ref{Eq:popdif}, as shown in Fig.~\ref{fig:fig3}(c). Through engineering the bath properties, the spin polarization could be preserved for 10s of ps.

We find spin-dependent electron transfer, when an achiral donor-acceptor dyad is excited with a circularly-polarized pump. Even in a randomly oriented ensemble of achiral donor–acceptor complexes, circularly polarized excitation defines a preferred direction $\mathbf{z}_{\mathrm{lab}}$ through its angular momentum, thereby preparing an oriented racemic mixture of transient enantiomers. As a result, when projected onto each molecule’s axis and then ensemble-averaged, the spin polarization vanishes, whereas the ensemble-averaged projection onto the laboratory axis is expected to be non-zero. This transient spin polarization can be measured using a spin-resolved photoemission probe at a time delay $\tau$ following the circularly polarized pump. By performing the measurement at multiple delays, provided their separation is not an integer multiple of the Rabi period $2\pi/\Omega$, the time-dependent spin polarization can be sampled.

Theoretical and experimental works in the 1960s-90s have established that photoelectrons ejected by circularly polarized light have spin angular momentum  correlated with the light's angular momentum due to spin-orbit coupling.\cite{fano1969spin,cherepkov1983spin} In such direct photoemission processes, the spin polarization arises during the optical transition itself, while the total spin of the combined system (ion + photoelectron) remains conserved. In contrast, in our case of photoinduced charge transfer with CPL, the optical transition prepares a singlet bound state with zero spin polarization. The spin polarization $\hat{P}_{z}$ becomes nonzero during electron transfer due to SOC and its value depends on the orbital angular momentum of the initial state. Here, CPL prepares a transiently chiral electronic state with orbital angular momentum, and spin polarization emerges dynamically through SOC-driven mixing between electronic states during charge transfer, rather than during the optical excitation itself. This suggests that our work is better understood as a transient-electronic CISS effect and is conceptually distinct from the generation of spin polarization during the optical transition, as occurs in photoemission with a circularly-polarized pump. Our mechanism is also related to light-induced ultrafast magnetism, where circularly polarized excitation generates transient spin polarization through spin-orbit coupling and symmetry breaking. Our results highlight a potential connection to these optically driven spin phenomena.\cite{okyay2020resonant,neufeld2025linearly}

In recent experimental work, Liu \textit{et al}. have demonstrated that upon exciting a transition metal dichalcogenide with CPL followed by electron transfer to an organic molecule, the generated spin polarization decays more slowly than in the bare monolayer.\cite{liu2024spin} However, in that work the spin of the electron in the conduction band is fixed at the time of excitation as up or down depending on the helicity of the light and this electron is then transferred due to spin-valley locking. Our proposal takes this a step further, as the CPL excitation in itself only generates a singlet and during the dynamics of electron transfer, the system develops spin polarization. This idea can be further developed to include other notions of transient nuclear chirality.\cite{zeng2025photo} The CISS effect has been demonstrated in donor-bridge-acceptor system when the bridge is chiral,\cite{eckvahl2023direct} and it is known that IR excitation of the bridge vibrational modes can modify electron transfer rates,\cite{delor2014toward} so in future work, it would be interesting to study spin-polarization during photoinduced electron transfer in a donor-bridge-acceptor system when an IR pulse excites vibrational modes making the bridge transiently chiral. Another promising future direction would be to use synthetic chiral light, which, unlike CPL, is chiral already within the electric-dipole approximation, potentially enabling enhanced spin polarization.\cite{ayuso2019synthetic,neufeld2020degree,mayer2022imprinting}

Acknowledgments.  We acknowledge funding from William M. Keck Foundation (8959) and  LDRD program at LBNL under the U.S. Department of Energy Office of Science, Office of Basic Energy Sciences, under Contract No. DEAC02-05CH11231.

\end{document}


\title{Supplementary information: Spin Polarization from Circularly Polarized Light Induced Charge Transfer}
\author{Sindhana Pannir-Sivajothi}
\affiliation{Department of Chemistry, University of California, Berkeley, California 94720, United States}
\author{David Limmer}
\email{dlimmer@berkeley.edu}
\affiliation{Department of Chemistry, University of California, Berkeley, California 94720, United States}
\affiliation{Kavli Energy Nanoscience Institute at Berkeley, Berkeley, California 94720, United States}
\affiliation{Chemical Sciences Division, Lawrence Berkeley National Laboratory, Berkeley, California 94720, United States}
\affiliation{Material Sciences Division, Lawrence Berkeley National Laboratory, Berkeley, California 94720, United States}
	\maketitle
\section{Gourterman four-orbital model}
Metalloporphyrins approximately belong to the $D_{4h}$ point group and the Gouterman orbitals include two exactly degenerate LUMOs that have $e_g$ symmetry and two accidentally degenerate HOMOs that have $a_{1u}$ and $a_{2u}$ symmetries \cite{gouterman1961spectra,gouterman1963spectra}. Let the HOMOs with $a_{2u}$ symmetry have creation operator $\hat{d}_{\mathrm{H}_1\sigma}^{\dagger}$ and with $a_{1u}$ symmetry have creation operator $\hat{d}_{\mathrm{H}_2\sigma}^{\dagger}$ where $\sigma=\alpha,\beta$ is the spin and we use restricted spin orbitals (the convention is the same as Gouterman's in \cite{gouterman1961spectra}). Let the LUMOs with $e_{gy}$ symmetry (transforms as $yz$) and $e_{gx}$ symmetry (transforms as $xz$) have creation operators $\hat{d}_{\mathrm{L}_1\sigma}^{\dagger}$ and $\hat{d}_{\mathrm{L}_2\sigma}^{\dagger}$, respectively. Then the transition dipole $\bra{\Psi_0}\hat{\boldsymbol{\mu}}\hat{d}_{\mathrm{L}_2\sigma}^{\dagger}\hat{d}_{\mathrm{H}_1\sigma}\ket{\Psi_0}$ is along $+x$, while $\bra{\Psi_0}\hat{\boldsymbol{\mu}}\hat{d}_{\mathrm{L}_1\sigma}^{\dagger}\hat{d}_{\mathrm{H}_2\sigma}\ket{\Psi_0}$ is along $-x$. On the other hand, both $\bra{\Psi_0}\hat{\boldsymbol{\mu}}\hat{d}_{\mathrm{L}_1\sigma}^{\dagger}\hat{d}_{\mathrm{H}_1\sigma}\ket{\Psi_0}$ and $\bra{\Psi_0}\hat{\boldsymbol{\mu}}\hat{d}_{\mathrm{L}_2\sigma}^{\dagger}\hat{d}_{\mathrm{H}_2\sigma}\ket{\Psi_0}$ are along $+y$. The singlet Q-band electronic states in terms of the real-valued frontier molecular orbitals with creation operators $\hat{d}_{\mathrm{H}_1\sigma}^{\dagger}$, $\hat{d}_{\mathrm{H}_2\sigma}^{\dagger}$, $\hat{d}_{\mathrm{L}_1\sigma}^{\dagger}$, and $\hat{d}_{\mathrm{L}_2\sigma}^{\dagger}$, are
\begin{subequations}
\begin{align}
    \ket{^1\mathrm{Q}_x}=&\frac{1}{2}\Big[(\hat{d}_{\mathrm{L}_2\alpha}^{\dagger}\hat{d}_{\mathrm{H}_1\alpha}+\hat{d}_{\mathrm{L}_2\beta}^{\dagger}\hat{d}_{\mathrm{H}_1\beta})+(\hat{d}_{\mathrm{L}_1\alpha}^{\dagger}\hat{d}_{\mathrm{H}_2\alpha}+\hat{d}_{\mathrm{L}_1\beta}^{\dagger}\hat{d}_{\mathrm{H}_2\beta})\Big]\ket{\Psi_0},\\
    \ket{^1\mathrm{Q}_y}=&\frac{1}{2}\Big[(\hat{d}_{\mathrm{L}_1\alpha}^{\dagger}\hat{d}_{\mathrm{H}_1\alpha}+\hat{d}_{\mathrm{L}_1\beta}^{\dagger}\hat{d}_{\mathrm{H}_1\beta})-(\hat{d}_{\mathrm{L}_2\alpha}^{\dagger}\hat{d}_{\mathrm{H}_2\alpha}+\hat{d}_{\mathrm{L}_2\beta}^{\dagger}\hat{d}_{\mathrm{H}_2\beta})\Big]\ket{\Psi_0}.
\end{align}
\end{subequations}
The donor $\lambda=+,-$ singlet and triplet excited states are defined as  \cite{barth2006unidirectional} 
\begin{subequations}
    \begin{align}
        \ket{^1\mathrm{D}_{\pm}}=&\frac{1}{\sqrt{2}}\Big(\ket{^1\mathrm{Q}_x}\pm i\ket{^1\mathrm{Q}_y}\Big)=\frac{1}{\sqrt{2}}(\hat{d}_{\mathrm{L}_{\pm}\alpha}^{\dagger}\hat{d}_{\mathrm{H}_{\pm}\alpha}+\hat{d}_{\mathrm{L}_{\pm}\beta}^{\dagger}\hat{d}_{\mathrm{H}_{\pm}\beta})\ket{\Psi_0},\\
        \ket{^3\mathrm{D}_{\pm}}=&\frac{1}{\sqrt{2}}\Big(\ket{^3\mathrm{Q}_x}\pm i\ket{^3\mathrm{Q}_y}\Big)=\frac{1}{\sqrt{2}}(\hat{d}_{\mathrm{L}_{\pm}\alpha}^{\dagger}\hat{d}_{\mathrm{H}_{\pm}\alpha}-\hat{d}_{\mathrm{L}_{\pm}\beta}^{\dagger}\hat{d}_{\mathrm{H}_{\pm}\beta})\ket{\Psi_0},
    \end{align}
\end{subequations}
where
\begin{subequations}\label{eq:LHlambda}
    \begin{align}
    \hat{d}_{\mathrm{L}_{\pm}\sigma}^{\dagger}=&\frac{1}{\sqrt{2}}(\hat{d}_{\mathrm{L}_2\sigma}^{\dagger}\pm i\hat{d}_{\mathrm{L}_1\sigma}^{\dagger}),\\
        \hat{d}_{\mathrm{H}_{\pm}\sigma}^{\dagger}=&\frac{1}{\sqrt{2}}(\hat{d}_{\mathrm{H}_1\sigma}^{\dagger}\pm i\hat{d}_{\mathrm{H}_2\sigma}^{\dagger}).
    \end{align}
\end{subequations}

\section{Molecular orbitals and H\"{u}ckel theory}
We write down a H\"{u}ckel Hamiltonian in the basis of the atomic orbitals involved in the $\pi$-conjugation within the donor and acceptor. 
Each donor atom contributes a $p_z$ orbital, each acceptor atom a $p_x$ orbital, and the metal center contributes both $d_{xz}$ and $d_{yz}$ orbitals. Let the donor-acceptor complex contain $N_a$ atoms, with indices $j \in \{1, 2, \dots, N_a-1\}$ labeling all atoms except the metal. Define $n_D$ as the set of donor atom indices and $n_A$ as those of the acceptor excluding the metal. Since only the metal is shared between donor and acceptor, $n_D \cap n_A = \emptyset$ and $n_D \cup n_A = \{1, 2, \dots, N_a - 1\}$. We denote the atomic orbitals (AOs) as $\ket{p_{z,j}}$ for $j \in n_D$, $\ket{p_{x,j}}$ for $j \in n_A$, and $\ket{d_{xz}}, \ket{d_{yz}}$ for the metal center. Let $nn_j$ denote the neighbors of atom $j$ (excluding the metal) and $nn_M$ the neighbors of the metal. The single-electron form of the mean-field Hamiltonian $\hat{h}=\hat{h}_{D}+\hat{h}_{A}+\hat{v}+\hat{v}_{\mathrm{SOC}}$ contains a H\"{u}ckel term that is partitioned into donor $\hat{h}_{\mathrm{D}}$, acceptor $\hat{h}_{\mathrm{A}}$, and coupling $\hat{v}$, and a mean-field SOC term $\hat{v}_{\mathrm{SOC}}$, where 
\begin{equation}
\begin{aligned}
    	\hat{h}_{\mathrm{D}}=&\sum_{j\in n_{D}} \alpha_{j}\ket{p_{z,j}}\bra{p_{z,j}}+\alpha_x\ket{d_{xz}}\bra{d_{xz}}+\alpha_y\ket{d_{yz}}\bra{d_{yz}}\\
        &+\sum_{j'\in nn_M\cap n_D}\Big(\beta_{xj'}\ket{d_{xz}}\bra{p_{z,j'}}+\beta_{yj'}\ket{d_{yz}}\bra{p_{z,j'}}+\mathrm{h.c.}\Big)+\sum_{j\in n_D}\sum_{j'\in nn_j}\beta_{jj'}\ket{p_{z,j}}\bra{p_{z,j'}},\\
        \hat{h}_{\mathrm{A}}=&\sum_{j\in n_{A}} \alpha_{j}\ket{p_{x,j}}\bra{p_{x,j}}+\alpha_x\ket{d_{xz}}\bra{d_{xz}}+\sum_{j\in n_A}\sum_{j'\in nn_j}\beta_{jj'}\ket{p_{x,j}}\bra{p_{x,j'}}+\sum_{j'\in nn_M\cap n_A}\Big(\beta_{xj'}\ket{d_{xz}}\bra{p_{x,j'}}+\mathrm{h.c.}\Big),\\
         \hat{v}=&-\alpha_x\ket{d_{xz}}\bra{d_{xz}},\\
 \hat{v}_{\mathrm{SOC}}=& \Big[ \xi_M\ket{d_{yz}}\bra{d_{xz}}i\hbar+\mathrm{h.c.}\Big]\cdot \hat{s}_{z},         
\end{aligned}
\end{equation}
with $\beta_{j'j}=\beta_{jj'}^*$, $\beta_{j'x}=\beta_{xj'}^*$ and $\beta_{j'y}=\beta_{yj'}^*$, and the term $\hat{v}=-\alpha_x\ket{d_{xz}}\bra{d_{xz}}$ accounts for the double counting of the $d_{xz}$ orbital energy in both $\hat{h}_{\mathrm{D}}$ and $\hat{h}_{\mathrm{A}}$.  The form $\hat{v}_{\mathrm{SOC}}$ of the SOC coupling can be derived from the Breit-Pauli form $\hat{H}_{\mathrm{SOC}}=\sum_j\sum_{I} \xi(r_{jI}) \hat{\mathbf{l}}_{jI} \cdot \hat{\mathbf{s}}_j$ where $j$ labels the electrons and $I$ the nuclei \cite{marian2012spin} keeping only single-center integrals. Within our AO basis, only the metal center contributes to the one-center SOC \cite{gouterman1973porphyrins,ake1969porphyrins}, $\hat{H}_{\mathrm{SOC}}\approx \sum_{j}\xi(r_{jM})\hat{l}_{jM,z_{\mathrm{mol}}} \hat{s}_{j,z_{\mathrm{mol}}}$ where $M$ labels the metal nucleus. The one-center SOC terms are defined as matrix elements where the angular momentum operator $\hat{\mathbf{l}}_{jI}$ and both AOs are centered around the same nucleus. Because the angular momentum operators are pure imaginary operators and have zero diagonal matrix elements in a real-valued AO basis and since we retain only one AO per atom except for the metal, it is evident that only the metal center will contribute to the one-center SOC. Therefore, we can write the single-electron SOC term of the Hamiltonian as $\hat{v}_{\mathrm{SOC}}=[\xi_M \ket{d_{yz}}\bra{d_{xz}}i\hbar+\mathrm{h.c.}]\hat{s}_{z}$ where $\xi_M=\bra{d_{yz}}\xi(r_{jM})\ket{d_{xz}}$ is real-valued.

By diagonalizing $\hat{h}_D$, we can compute the degenerate pair of HOMOs and LUMOs of the donor given by creation operators $\hat{d}^{\dagger}_{\mathrm{H}_{1}\sigma}$, $\hat{d}^{\dagger}_{\mathrm{H}_{2}\sigma}$, $\hat{d}^{\dagger}_{\mathrm{L}_{1}\sigma}$ and $\hat{d}^{\dagger}_{\mathrm{L}_{2}\sigma}$ as a LCAO, and similarly by diagonalizing $\hat{h}_{A}$, we can obtain the LUMO of the acceptor given by creation operator $\hat{a}^{\dagger}_{\mathrm{L\sigma}}$. The MOs are $\hat{d}_{\mathrm{L}_{\lambda}\sigma}^{\dagger}\ket{0}=\ket{\phi_{\mathrm{D},\mathrm{L}_{\lambda}}}\ket{\sigma}$, $\hat{d}_{\mathrm{H}_{\lambda}\sigma}^{\dagger}\ket{0}=\ket{\phi_{\mathrm{D},\mathrm{H}_{\lambda}}}\ket{\sigma}$, $\hat{a}_{\mathrm{L}\sigma}^{\dagger}\ket{0}=\ket{\phi_{\mathrm{A},\mathrm{L}}}\ket{\sigma}$, and they satisfy $\hat{h}_{\mathrm{D}}\ket{\phi_{\mathrm{D},\mathrm{L}_{\lambda}}}=\epsilon_{\mathrm{D,L}}\ket{\phi_{\mathrm{D},\mathrm{L}_{\lambda}}}$, $\hat{h}_{\mathrm{D}}\ket{\phi_{\mathrm{D},\mathrm{H}_{\lambda}}}=\epsilon_{\mathrm{D,H}}\ket{\phi_{\mathrm{D},\mathrm{H}_{\lambda}}}$, and $\hat{h}_{\mathrm{A}}\ket{\phi_{\mathrm{A},\mathrm{L}}}=\epsilon_{\mathrm{A,L}}\ket{\phi_{\mathrm{A,L}}}$. Both the spin-preserving and the spin-orbit coupled electron transfer happens via the $d_{\pi}$ orbitals of the metal; therefore, the contribution of the $d_{\pi}$ orbitals of the metal to the LUMOs of the donor and acceptor determine the charge transfer matrix elements. Due to symmetry, the $d_{xz}$ orbital only contributes to $\hat{d}_{\mathrm{L}_2\sigma}^{\dagger}$ and $\hat{a}_{\mathrm{L}\sigma}^{\dagger}$ MOs, and $d_{yz}$ orbitals only to $\hat{d}_{\mathrm{L}_1\sigma}^{\dagger}$. Since $\hat{d}_{\mathrm{L}_1\sigma}^{\dagger}$, $\hat{d}_{\mathrm{L}_2\sigma}^{\dagger}$ and $\hat{a}_{\mathrm{L}\sigma}^{\dagger}$ are creation operators for purely real MOs, $B_{xz}=\bra{d_{xz}}\ket{\phi_{\mathrm{D},\mathrm{L}_{2}}}$, $B_{yz}=\bra{d_{yz}}\ket{\phi_{\mathrm{D},\mathrm{L}_{1}}}$ and $A_{xz}=\bra{d_{xz}}\ket{\phi_{\mathrm{A},\mathrm{L}}}$ are purely real. Then we can show using Eq. \ref{eq:LHlambda} that $\bra{d_{xz}}\ket{\phi_{\mathrm{D},\mathrm{L}_{\pm}}}=B_{xz}/\sqrt{2}$ and $\bra{d_{yz}}\ket{\phi_{\mathrm{D},\mathrm{L}_{\pm}}}=\pm i B_{yz}/\sqrt{2}$. The matrix elements of the frontier MOs are
\begin{equation}\label{eq:matrixelement1}
	\begin{aligned}
\bra{\phi_{\mathrm{D,H}_{\lambda}}}(\hat{h}_{\mathrm{D}}+\hat{h}_{\mathrm{A}}+\hat{v})\ket{\phi_{\mathrm{D,H}_{\lambda}}}=&\epsilon_{\mathrm{D,H}}\bra{\phi_{\mathrm{D,H}_{\lambda}}}\ket{\phi_{\mathrm{D,H}_{\lambda}}}+\bra{\phi_{\mathrm{D,H}_{\lambda}}}(\hat{h}_{A}+\hat{v})\ket{\phi_{\mathrm{D,H}_{\lambda}}}\\
		=&\epsilon_{\mathrm{D,H}}\\
\bra{\phi_{\mathrm{D,L}_{\lambda}}}(\hat{h}_{\mathrm{D}}+\hat{h}_{\mathrm{A}}+\hat{v})\ket{\phi_{\mathrm{D,L}_{\lambda}}}=&\epsilon_{\mathrm{D,L}}\bra{\phi_{\mathrm{D,L}_{\lambda}}}\ket{\phi_{\mathrm{D,L}_{\lambda}}}+\bra{\phi_{\mathrm{D,L}_{\lambda}}}(\hat{h}_{A}+\hat{v})\ket{\phi_{\mathrm{D,L}_{\lambda}}}\\
\approx&\epsilon_{\mathrm{D,L}}\\
\bra{\phi_{\mathrm{A,L}}}(\hat{h}_{\mathrm{D}}+\hat{h}_{\mathrm{A}}+\hat{v})\ket{\phi_{\mathrm{A,L}}}=&\epsilon_{\mathrm{A,L}}\bra{\phi_{\mathrm{A,L}}}\ket{\phi_{\mathrm{A,L}}}+\bra{\phi_{\mathrm{A,L}}}(\hat{h}_D+\hat{v})\ket{\phi_{\mathrm{A,L}}}\\
\approx&\epsilon_{\mathrm{A,L}}\\
\bra{\phi_{\mathrm{D,L}_{\lambda}}}(\hat{h}_{\mathrm{D}}+\hat{h}_{\mathrm{A}}+\hat{v})\ket{\phi_{\mathrm{A,L}}}=&\bra{\phi_{\mathrm{D,L}_{\lambda}}}\hat{h}_{\mathrm{D}}\ket{\phi_{\mathrm{A,L}}}+\bra{\phi_{\mathrm{D,L}_{\lambda}}}\hat{h}_{\mathrm{A}}\ket{\phi_{\mathrm{A,L}}}+\bra{\phi_{\mathrm{D,L}_{\lambda}}}\hat{v}\ket{\phi_{\mathrm{A,L}}}\\
		=&\epsilon_{\mathrm{D,L}}\bra{\phi_{\mathrm{D,L}_{\lambda}}}\ket{\phi_{\mathrm{A,L}}}+\epsilon_{\mathrm{A,L}}\bra{\phi_{\mathrm{D,L}_{\lambda}}}\ket{\phi_{\mathrm{A,L}}}-\alpha_x\bra{\phi_{\mathrm{D,L}_{\lambda}}}\ket{d_{xz}}\bra{d_{xz}}\ket{\phi_{\mathrm{A,L}}}\\
		\approx&(\epsilon_{\mathrm{D,L}}+\epsilon_{\mathrm{A,L}}-\alpha_x)\bra{\phi_{\mathrm{D,L}_{\lambda}}}\ket{d_{xz}}\bra{d_{xz}}\ket{\phi_{\mathrm{A,L}}}\\
		=&(\epsilon_{\mathrm{D,L}}+\epsilon_{\mathrm{A,L}}-\alpha_x)B_{xz}A_{xz}/\sqrt{2}
	\end{aligned}
\end{equation}
and
\begin{equation}\label{eq:matrixelement2}
    \begin{aligned}
            \bra{\phi_{\mathrm{D,H}_{\lambda}}}(\hat{h}_{\mathrm{D}}+\hat{h}_{\mathrm{A}}+\hat{v})\ket{\phi_{\mathrm{A,L}}}=&\bra{\phi_{\mathrm{D,H}_{\lambda}}}\hat{h}_{\mathrm{D}}\ket{\phi_{\mathrm{A,L}}}+\bra{\phi_{\mathrm{D,H}_{\lambda}}}\hat{h}_{\mathrm{A}}\ket{\phi_{\mathrm{A,L}}}+\bra{\phi_{\mathrm{D,H}_{\lambda}}}\hat{v}\ket{\phi_{\mathrm{A,L}}}\\
		=&\epsilon_{\mathrm{D,H}}\bra{\phi_{\mathrm{D,H}_{\lambda}}}\ket{\phi_{\mathrm{A,L}}}+\epsilon_{\mathrm{A,L}}\bra{\phi_{\mathrm{D,H}_{\lambda}}}\ket{\phi_{\mathrm{A,L}}}-\alpha_x\bra{\phi_{\mathrm{D,H}_{\lambda}}}\ket{d_{xz}}\bra{d_{xz}}\ket{\phi_{\mathrm{A,L}}}\\
		\approx&(\epsilon_{\mathrm{D,H}}+\epsilon_{\mathrm{A,L}}-\alpha_x)\bra{\phi_{\mathrm{D,H}_{\lambda}}}\ket{d_{xz}}\bra{d_{xz}}\ket{\phi_{\mathrm{A,L}}}\\
				=&0\\
		\bra{\phi_{\mathrm{D,H}_{\lambda_1}}}(\hat{h}_{\mathrm{D}}+\hat{h}_{\mathrm{A}}+\hat{v})\ket{\phi_{\mathrm{D,L}_{\lambda_2}}}=&0\\
        \bra{\phi_{\mathrm{D,H}_{+}}}(\hat{h}_{\mathrm{D}}+\hat{h}_{\mathrm{A}}+\hat{v})\ket{\phi_{\mathrm{D,H}_{-}}}=&0\\
        \bra{\phi_{\mathrm{D,L}_{+}}}(\hat{h}_{\mathrm{D}}+\hat{h}_{\mathrm{A}}+\hat{v})\ket{\phi_{\mathrm{D,L}_{-}}}=&\bra{\phi_{\mathrm{D,L}_{+}}}(\hat{h}_{A}+\hat{v})\ket{\phi_{\mathrm{D,L}_{-}}}\\
        \approx&0\\
\bra{\phi_{\mathrm{D,L}_{\pm}}}\bra{\sigma_1}\hat{v}_{\mathrm{SOC}}\ket{\sigma_2}\ket{\phi_{\mathrm{D,L}_{\pm}}}=&\bra{\phi_{\mathrm{D,L}_{\pm}}}\Big[ i\hbar\xi_M\ket{d_{yz}}\bra{d_{xz}}-i\hbar\xi_{\mathrm{M}}\ket{d_{xz}}\bra{d_{yz}}\Big]\ket{\phi_{\mathrm{D,L}_{\pm}}}\bra{\sigma_1}\hat{s}_{z}\ket{\sigma_2}\\
=&\Big[ i\hbar\xi_M(\mp i B_{yz})B_{xz}/2-i\hbar\xi_MB_{xz}(\pm iB_{yz}/2)\Big]\bra{\sigma_1}\hat{s}_{z}\ket{\sigma_2}\\
=&\pm\bra{\sigma_1}\hat{s}_{z}\ket{\sigma_2}\hbar\xi_M B_{yz}B_{xz}\\
\bra{\phi_{\mathrm{D,L}_{+}}}\bra{\sigma_1}\hat{v}_{\mathrm{SOC}}\ket{\sigma_2}\ket{\phi_{\mathrm{D,L}_{-}}}=&\bra{\phi_{\mathrm{D,L}_{+}}}\Big[ i\hbar\xi_M\ket{d_{yz}}\bra{d_{xz}}-i\hbar\xi_{\mathrm{M}}\ket{d_{xz}}\bra{d_{yz}}\Big]\ket{\phi_{\mathrm{D,L}_{-}}}\bra{\sigma_1}\hat{s}_{z}\ket{\sigma_2}\\
=&\Big[ i\hbar\xi_M(\mp i B_{yz})B_{xz}/2-i\hbar\xi_MB_{xz}(\mp iB_{yz}/2)\Big]\bra{\sigma_1}\hat{s}_{z}\ket{\sigma_2}\\
=&0\\
\bra{\phi_{\mathrm{D,H}_{\lambda_1}}}\bra{\sigma_1}\hat{v}_{\mathrm{SOC}}\ket{\sigma_2}\ket{\phi_{\mathrm{D,H}_{\lambda_2}}}=&0\\
\bra{\phi_{\mathrm{D,L}_{\pm}}}\bra{\sigma_1}\hat{v}_{\mathrm{SOC}}\ket{\sigma_2}\ket{\phi_{\mathrm{A,L}}}=&\bra{\phi_{L_{\pm}}} \Big[ i\hbar\xi_M\ket{d_{yz}}\bra{d_{xz}}-i\hbar\xi_M\ket{d_{xz}}\bra{d_{yz}}\Big]\ket{\phi_{L,a}}\bra{\sigma_1}\hat{s}_{z}\ket{\sigma_2}\\
=&  \pm\bra{\sigma_1}\hat{s}_{z}\ket{\sigma_2}\hbar\xi_MB_{yz}A_{xz}/\sqrt{2}\\
\bra{\phi_{\mathrm{A,L}}}\bra{\sigma_1}\hat{v}_{\mathrm{SOC}}\ket{\sigma_2}\ket{\phi_{\mathrm{A,L}}}=&0
    \end{aligned}
\end{equation}
where $\lambda_1,\lambda_2\in\{+,-\}$ and we have assumed that the overlap of atomic orbitals of non-neighboring atoms is zero, so $\bra{p_{z,j}}\ket{p_{x,j'}}=0$. The atomic orbital basis is non-orthogonal; therefore, the orbital $\ket{d_{xz}}$ has non-zero overlap with $\ket{p_{z,j}}$ of adjacent atoms, so $\bra{\phi_{\mathrm{D,L}_{\pm}}}\ket{\phi_{\mathrm{A,L}}}\neq \bra{\phi_{\mathrm{D,L}_{\pm}}}\ket{d_{xz}}\bra{d_{xz}}\ket{\phi_{\mathrm{A,L}}}$ but it will be almost equal to it.

The many-electron mean-field Hamiltonian is $\hat{H}_{\mathrm{MF}}=\sum_j\Big[\hat{h}_{\mathrm{D}}(j)+\hat{h}_{\mathrm{A}}(j)+\hat{v}(j)+\hat{v}_{\mathrm{SOC}}(j)\Big]$ where $j$ labels the electrons. After projecting this Hamiltonian into the subspace of frontier orbitals using the matrix elements in Eq. \ref{eq:matrixelement1} and \ref{eq:matrixelement2}, it decomposes into $\lambda=\pm$ sectors $\hat{\Pi}_{\mathrm{frontier}}\hat{H}_{\mathrm{MF}}\hat{\Pi}_{\mathrm{frontier}}=\sum_{\lambda=+,-}\hat{H}_{\lambda}$ with
\begin{equation}\label{eq:HamilFrontierSI}
\begin{aligned}  \hat{H}_{\lambda}=&\sum_{\sigma=\alpha,\beta}\Big(\epsilon_{\mathrm{D,L}}\hat{d}^{\dagger}_{\mathrm{L}_{\lambda}\sigma}\hat{d}_{\mathrm{L}_{\lambda}\sigma}+\epsilon_{\mathrm{D,H}}\hat{d}^{\dagger}_{\mathrm{H}_{\lambda}\sigma}\hat{d}_{\mathrm{H}_{\lambda}\sigma}+\epsilon_{\mathrm{A,L}}\hat{a}_{\mathrm{L}\sigma}^{\dagger}\hat{a}_{\mathrm{L}\sigma}+v_0\hat{d}^{\dagger}_{\mathrm{L}_{\lambda}\sigma}\hat{a}_{\mathrm{L}\sigma}+v_0\hat{d}_{\mathrm{L}_{\lambda}\sigma}\hat{a}^{\dagger}_{\mathrm{L}\sigma}\Big)\\
&+v_{\mathrm{ISC}_{\lambda}}\Big(\hat{d}_{\mathrm{L}_{\lambda}\alpha}^{\dagger}\hat{d}_{\mathrm{L}_{\lambda}\alpha}-\hat{d}_{\mathrm{L}_{\lambda}\beta}^{\dagger}\hat{d}_{\mathrm{L}_{\lambda}\beta}\Big)+v_{\mathrm{SOC}_{\lambda}}\Big(\hat{d}_{\mathrm{L}_{\lambda}\alpha}^{\dagger}\hat{a}_{\mathrm{L}\alpha} -\hat{d}_{\mathrm{L}_{\lambda}\beta}^{\dagger}\hat{a}_{\mathrm{L}\beta}+\mathrm{h.c.}\Big),
\end{aligned}
\end{equation}
where $v_0\approx(\epsilon_{\mathrm{D,L}}+\epsilon_{\mathrm{A,L}}-\alpha_x) A_{xz} B_{xz}/\sqrt{2}$, $v_{\mathrm{ISC}_{\pm}}=\pm\xi_M B_{xz}B_{yz}\hbar^2/2$ and $v_{\mathrm{SOC}_{\pm}}=\pm\xi_M A_{xz}B_{yz}\hbar^2/2\sqrt{2}$ are all real-valued.

The mean-field Hamiltonian $\hat{H}_{\mathrm{MF}}$ only contains single-electron terms and does not include exchange coupling that results in the singlet-triplet energy gap. We can phenomenologically include the two-electron exchange interaction term $\hat{V}_{\mathrm{ex}}=J\sum_{\lambda_1}\sum_{\lambda_2}\sum_{\sigma=\alpha,\beta}\sum_{\tau=\alpha,\beta}\hat{d}^{\dagger}_{\mathrm{L}_{\lambda_2}\sigma}\hat{d}^{\dagger}_{\mathrm{H}_{\lambda_1}\tau}\hat{d}_{\mathrm{L}_{\lambda_2}\tau}\hat{d}_{\mathrm{H}_{\lambda_1}\sigma}$ that results in an energy gap between $\ket{^1\mathrm{D}_{\lambda}}$ and $\ket{^3\mathrm{D}_{\lambda}}$. The Hamiltonian $\hat{H}_{\mathrm{e}}$ defined in Eq. 3 of the main text represents $\hat{H}_{\mathrm{MF}}+\hat{V}_{\mathrm{ex}}$ in the subspace spanned by $\ket{^1\mathrm{D}_{\lambda}}$, $\ket{^1\mathrm{CT}_{\lambda}}$, and $\ket{^3\mathrm{CT}_{\lambda}}$ as the triplet donor is far detuned from the other states, and therefore all calculations in this work are performed with this Hamiltonian. For completeness, here we include the mean-field Hamiltonian along with the exchange coupling in the space spanned by $\ket{^1\mathrm{D}_{\lambda}}$, $\ket{^3\mathrm{D}_{\lambda}}$, $\ket{^1\mathrm{CT}_{\lambda}}$, and $\ket{^3\mathrm{CT}_{\lambda}}$, 
\begin{equation}\label{eq:Helec2}
\begin{aligned}  \hat{H}_{\mathrm{e}-4}=&\sum_{\lambda=+,-}\Big(\Delta\ket{^1\mathrm{CT}_{\lambda}}\bra{^1\mathrm{CT}_{\lambda}}+\Delta\ket{^3\mathrm{CT}_{\lambda}}\bra{^3\mathrm{CT}_{\lambda}}-2J\ket{^3\mathrm{D}_{\lambda}}\bra{^3\mathrm{D}_{\lambda}}\\
&+v_0\ket{^1\mathrm{D}_{\lambda}}\bra{^1\mathrm{CT}_{\lambda}}+v_0\ket{^1\mathrm{CT}_{\lambda}}\bra{^1\mathrm{D}_{\lambda}}+v_0\ket{^3\mathrm{D}_{\lambda}}\bra{^3\mathrm{CT}_{\lambda}}+v_0\ket{^3\mathrm{CT}_{\lambda}}\bra{^3\mathrm{D}_{\lambda}}\\
&+v_{\mathrm{SOC}_{\lambda}}\ket{^1\mathrm{D}_{\lambda}}\bra{^3\mathrm{CT}_{\lambda}}+v_{\mathrm{SOC}_{\lambda}}\ket{^3\mathrm{CT}_{\lambda}}\bra{^1\mathrm{D}_{\lambda}}+v_{\mathrm{SOC}_{\lambda}}\ket{^3\mathrm{D}_{\lambda}}\bra{^1\mathrm{CT}_{\lambda}}+v_{\mathrm{SOC}_{\lambda}}\ket{^1\mathrm{CT}_{\lambda}}\bra{^3\mathrm{D}_{\lambda}}\\
&+v_{\mathrm{ISC}_{\lambda}}\ket{^1\mathrm{D}_{\lambda}}\bra{^3\mathrm{D}_{\lambda}}+v_{\mathrm{ISC}_{\lambda}}\ket{^3\mathrm{D}_{\lambda}}\bra{^1\mathrm{D}_{\lambda}}\Big).
\end{aligned}
\end{equation}
 We omit the triplet states with spin projection $M_S=\pm 1$, because the initial state $\ket{^1\mathrm{D}_{\pm}}$ lies in the $M_S=0$ manifold, and given $[\sum_{\lambda}\hat{H}_{\lambda},\hat{S}_z]=0$, the dynamics conserves $M_S$.

The term with coefficient $v_{\mathrm{ISC}_{\lambda}}$ in Eq. \ref{eq:Helec2} comes from the difference in energy between a spin-up and a spin-down electron occupying the $\hat{d}^{\dagger}_{\mathrm{L}_{\pm},\sigma}$ orbitals due to the magnetic field generated by the ring currents in these MOs. The orbital angular momentum of these states interacts with the spin analogous to an effective magnetic field via spin-orbit coupling. By comparison, due to the magnetic field generated by the hole current in the charge transfer states (which is why they retain the label $\lambda$), we would expect intersystem crossing between charge transfer states $\ket{^1\mathrm{CT}_{\lambda}}$ and $\ket{^3\mathrm{CT}_{\lambda}}$. However, since the HOMOs of the donor do not have any metal $d_{xz}$ or $d_{yz}$ character, they do not contribute to the spin-orbit coupling and therefore any orbital angular momentum of the hole does not produce an effective magnetic field for the electron spin in the charge transfer states.

The z-component of electronic orbital angular momentum $\hat{L}_{z_{\mathrm{mol}}}=\sum_j \hat{\mathbf{l}}_{jM}\cdot\mathbf{z}_{\mathrm{mol}}=\sum_j \hat{l}_{jM,z}$ where $\hat{\mathbf{l}}_{jM}$ denotes the orbital angular momentum of electron $j$ about the metal center $M$. We assume the frontier molecular orbitals of the donor to be eigenstates of $\hat{l}_{jM,z}$,
\begin{subequations}
    \begin{equation}
\hat{l}_{jM,z}\ket{\phi_{\mathrm{D,H}_{\lambda}}}=\lambda \hbar m_{\mathrm{HOMO}}\ket{\phi_{\mathrm{D,H}_{\lambda}}},
    \end{equation}
        \begin{equation}
\hat{l}_{jM,z}\ket{\phi_{\mathrm{D,L}_{\lambda}}}=\lambda \hbar m_{\mathrm{LUMO}}\ket{\phi_{\mathrm{D,L}_{\lambda}}}.
    \end{equation}
\end{subequations}
This assumption is well justified for porphyrins, whose $\pi$-system can be approximated as a cyclic polyene within a H\"{u}ckel description \cite{perrin1969vibronic}. In that limit, the eigenstates are discrete Bloch waves on a ring, which serve as discrete analogues of angular momentum eigenstates of a particle on a ring \cite{albright2013orbital}. The frontier orbitals of the donor form degenerate pairs corresponding to clockwise, $\lambda=-$, and counterclockwise, $\lambda=+$, ring currents. The acceptor LUMO is a purely real orbital, therefore $\bra{\phi_{\mathrm{A,L}}}\hat{l}_{jM,z}\ket{\phi_{\mathrm{A,L}}}=0$. Using these matrix elements, in second-quantized notation, the angular momentum operator
\begin{equation}\label{eq:Lz}
    \hat{L}_{z_{\mathrm{mol}}}=\sum_{\lambda=+,-}\sum_{\sigma=\alpha,\beta}\lambda \hbar m_{\mathrm{LUMO}}\hat{d}^{\dagger}_{\mathrm{L}_{\lambda}\sigma}\hat{d}_{\mathrm{L}_{\lambda}\sigma}+\lambda \hbar m_{\mathrm{HOMO}}\hat{d}^{\dagger}_{\mathrm{H}_{\lambda}\sigma}\hat{d}_{\mathrm{H}_{\lambda}\sigma}+\Big(\lambda \hbar m_{\mathrm{LUMO}}B_{xz}A_{xz}\hat{a}^{\dagger}_{\mathrm{L}\sigma}\hat{d}_{\mathrm{L}_{\lambda}\sigma}+\mathrm{H.c.}\Big).
\end{equation}
Here, the coefficient of the $\hat{a}^{\dagger}_{\mathrm{L}\sigma}\hat{d}_{\mathrm{H}_{\lambda}\sigma}$ term is almost zero as the donor HOMO does not have any $d_{xz}$ character and $\bra{\phi_{\mathrm{A,L}}}\ket{\phi_{\mathrm{D,H}_{\lambda}}}\approx 0$, so we don't include it in Eq. \ref{eq:Lz}. The representation of this angular momentum in the basis of electronic states is
\begin{equation}
    \begin{aligned}
\hat{L}_{z_{\mathrm{mol}}}=&\sum_{\lambda=+,-}\Big[\lambda\hbar(m_{\mathrm{LUMO}}-m_{\mathrm{HOMO}})\ket{^1\mathrm{D}_{\lambda}}\bra{^1\mathrm{D}_{\lambda}}-\lambda\hbar m_{\mathrm{HOMO}}\ket{^1\mathrm{CT}_{\lambda}}\bra{^1\mathrm{CT}_{\lambda}}-\lambda \hbar m_{\mathrm{HOMO}}\ket{^3\mathrm{CT}_{\lambda}}\bra{^3\mathrm{CT}_{\lambda}}\\
&+\lambda\hbar m_{\mathrm{LUMO}} B_{xz}A_{xz}\Big(\ket{^1\mathrm{D}_{\lambda}}\bra{^1\mathrm{CT}_{\lambda}}+\ket{^1\mathrm{D}_{\lambda}}\bra{^3\mathrm{CT}_{\lambda}}+\mathrm{H.c.}\Big)\Big].
    \end{aligned}
\end{equation}
The $d_{xz}$ component of the donor LUMO, $B_{xz}$, and that of the acceptor LUMO, $A_{xz}$, will be smaller than $1$. If $A_{xz}, B_{xz}\ll 1$, then the electronic states are approximate eigenstates of the angular momentum operator
\begin{subequations}
\begin{align}
\hat{L}_{z_{\mathrm{mol}}}\ket{^1\mathrm{D}_{\lambda}}\approx&\lambda\hbar(m_{\mathrm{LUMO}}-m_{\mathrm{HOMO}})\ket{^1\mathrm{D}_{\lambda}},\\     \hat{L}_{z_{\mathrm{mol}}}\ket{^{1(3)}\mathrm{CT}_{\lambda}}\approx&-\lambda\hbar m_{\mathrm{HOMO}}\ket{^{1(3)}\mathrm{CT}_{\lambda}}.
\end{align}
\end{subequations}

\section{Dynamics without vibronic coupling}
In this section, we solve for the dynamics of the system under unitary evolution with $\hat{H}_{\mathrm{e}}$, in the absence of vibronic coupling. As the charge transfer states are degenerate, we change basis such that only one of them couples to the singlet donor $\ket{^1\mathrm{D}_{\lambda}}$. These new charge transfer states include one that has mostly singlet character 
$\ket{^{1\gg 3}\mathrm{CT}_{\lambda}}$ and one with mostly triplet character $\ket{^{3\gg 1}\mathrm{CT}_{\lambda}}$,
\begin{subequations}
    \begin{align}
        \ket{^{1\gg 3}\mathrm{CT}_{\lambda}}=&\frac{v_0}{\sqrt{v_0^2+v_{\mathrm{SOC}_{\lambda}}^2}}\ket{^1\mathrm{CT}_{\lambda}}+\frac{v_{\mathrm{SOC}_{\lambda}}}{\sqrt{v_0^2+v_{\mathrm{SOC}_{\lambda}}^2}}\ket{^3\mathrm{CT}_{\lambda}}\\
        \ket{^{3\gg 1}\mathrm{CT}_{\lambda}}=&-\frac{v_{\mathrm{SOC}_{\lambda}}}{\sqrt{v_0^2+v_{\mathrm{SOC}_{\lambda}}^2}}\ket{^1\mathrm{CT}_{\lambda}}+\frac{v_0}{\sqrt{v_0^2+v_{\mathrm{SOC}_{\lambda}}^2}}\ket{^3\mathrm{CT}_{\lambda}}
    \end{align}
\end{subequations}
In this basis, the Hamiltonian
\begin{equation}
\begin{aligned}
\hat{H}_{\mathrm{e}}=&\sum_{\lambda=+,-}\Bigg[\Delta\ket{^{1\gg 3}\mathrm{CT}_{\lambda}}\bra{^{1\gg 3}\mathrm{CT}_{\lambda}}+\Delta\ket{^{3\gg 1}\mathrm{CT}_{\lambda}}\bra{^{3\gg 1}\mathrm{CT}_{\lambda}}\\
&+\sqrt{v_0^2+v_{\mathrm{SOC}_{\lambda}}^2}\Big(\ket{^{1\gg 3}\mathrm{CT}_{\lambda}}\bra{^1\mathrm{D}_{\lambda}}+\ket{^1\mathrm{D}_{\lambda}}\bra{^{1\gg 3}\mathrm{CT}_{\lambda}}\Big)\Bigg].
\end{aligned}
\end{equation}
The spin polarization in this basis is
\begin{equation}
 \begin{aligned}
\hat{P}_{z}=&\sum_{\lambda}\frac{\hbar}{v_0^2+v_{\mathrm{SOC}_{\lambda}}^2}\Bigg[2v_0v_{\mathrm{SOC}_{\lambda}}\ket{^{1\gg 3}\mathrm{CT}_{\lambda}}\bra{^{1\gg 3}\mathrm{CT}_{\lambda}}-2v_0v_{\mathrm{SOC}_{\lambda}}\ket{^{3\gg 1}\mathrm{CT}_{\lambda}}\bra{^{3\gg 1}\mathrm{CT}_{\lambda}}\\
 &+(v_0^2-v_{\mathrm{SOC}_{\lambda}}^2)\Big(\ket{^{1\gg 3}\mathrm{CT}_{\lambda}}\bra{^{3\gg 1}\mathrm{CT}_{\lambda}}+\mathrm{h.c.}\Big)\Bigg].
 \end{aligned}
\end{equation}
The eigenstates of $\hat{H}_{\mathrm{e}}$ are
\begin{subequations}
\begin{align}
    \ket{^1+_{\lambda}}=&\cos\theta \ket{^1\mathrm{D}_{\lambda}}+\sin\theta\ket{^{1\gg 3}\mathrm{CT}_{\lambda}},\\
    \ket{^1-_{\lambda}}=&-\sin\theta \ket{^1\mathrm{D}_{\lambda}}+\cos\theta\ket{^{1\gg 3}\mathrm{CT}_{\lambda}},\\
    \ket{^{3\gg 1}\mathrm{CT}_{\lambda}}=&\ket{^{3\gg 1}\mathrm{CT}_{\lambda}},
\end{align}
\end{subequations}
with energies $E_{1+}=\Delta/2+\Omega$, $E_{1-}=\Delta/2-\Omega$ and $E_{3}=\Delta$,  where $\Omega=\sqrt{(\Delta/2)^2+v_0^2+v_{\mathrm{SOC}_{\lambda}}^2}$, $\cos\theta=\sqrt{(\Omega-\Delta/2)/2\Omega}$, and $\sin\theta=\sqrt{(\Omega+\Delta/2)/2\Omega}$. When the initial state is $\ket{^1\mathrm{D}_{\lambda}}$, the donor singlet population is
\begin{equation}
\begin{aligned}
p_{\ket{^1\mathrm{D}_+}}(t)=&\bra{^1\mathrm{D}_{\lambda}}e^{i\hat{H}_{\mathrm{e}}t/\hbar}\ket{^1\mathrm{D}_{\lambda}}\bra{^1\mathrm{D}_{\lambda}}e^{-i\hat{H}_{\mathrm{e}}t/\hbar}\ket{^1\mathrm{D}_{\lambda}}\\
=&\Big(e^{iE_{1+}t/\hbar}\cos\theta\bra{^1+_{\lambda}}-e^{iE_{1-}t/\hbar}\sin\theta\bra{^1-_{\lambda}}\Big)\ket{^1\mathrm{D}_{\lambda}}\bra{^1\mathrm{D}_{\lambda}}\\
  &\Big(e^{-iE_{1+}t/\hbar}\cos\theta\ket{^1+_{\lambda}}-e^{-iE_{1-}t/\hbar}\sin\theta\ket{^1-_{\lambda}}\Big)\\
  =&1-\frac{v_0^2+v_{\mathrm{SOC}_{\lambda}}^2}{2\Omega^2}\Big[1-\cos(2\Omega t/\hbar)\Big].
\end{aligned}
\end{equation}
Similarly, when the donor singlet is the initial state, the charge-transfer states have populations
\begin{subequations}
    \begin{equation}
p_{\ket{^1\mathrm{CT}_+}}(t)=\frac{v_0^2}{2\Omega^2}\Big[1-\cos(2\Omega t/\hbar)\Big],
\end{equation}
\begin{equation}
p_{\ket{^3\mathrm{CT}_+}}(t)=\frac{v_{\mathrm{SOC}_{+}}^2}{2\Omega^2}\Big[1-\cos(2\Omega t/\hbar)\Big].
\end{equation}
\end{subequations}
When the initial state is  $\ket{^1\mathrm{D}_{\lambda}}$, the spin polarization
\begin{equation}
  \begin{aligned}
  \langle \hat{P}_{z}(t)\rangle=&\bra{^1\mathrm{D}_{\lambda}}e^{i\hat{H}_{\mathrm{e}}t/\hbar}\delta \hat{P}_{z_{\mathrm{mol}}}e^{-i\hat{H}_{\mathrm{e}}t/\hbar}\ket{^1\mathrm{D}_{\lambda}}\\
=&\Big(e^{iE_{1+}t/\hbar}\cos\theta\bra{^1+_{\lambda}}-e^{iE_{1-}t/\hbar}\sin\theta\bra{^1-_{\lambda}}\Big)\frac{2\hbar v_0v_{\mathrm{SOC}_{\lambda}}}{v_0^2+v_{\mathrm{SOC}_{\lambda}}^2}\ket{^{1\gg 3}\mathrm{CT}}\bra{^{1\gg 3}\mathrm{CT}}\\
  &\Big(e^{-iE_{1+}t/\hbar}\cos\theta\ket{^1+_{\lambda}}-e^{-iE_{1-}t/\hbar}\sin\theta\ket{^1-_{\lambda}}\Big)\\
  =&\frac{\hbar v_0v_{\mathrm{SOC}_{\lambda}}}{\Omega^2}\Big[1-\cos(2\Omega t/\hbar)\Big].
  \end{aligned}
\end{equation}

\section{Relaxation rates: Redfield tensor}
\begin{figure*}
	\includegraphics[width=0.5\linewidth]{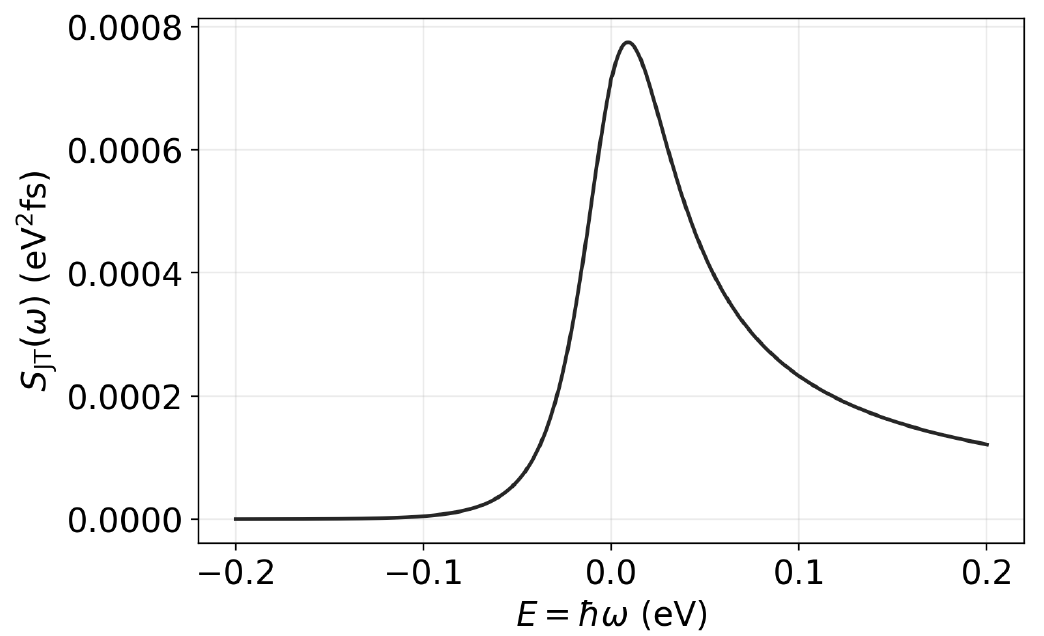}
	\caption{\label{fig:figS1} The real part of the spectral function $S_{\mathrm{JT}}(\omega)=\Re[\tilde{C}(\omega)]=\pi \hbar J(|\omega|)\Big[n(|\omega|)+\Theta(\omega)\Big]$ for a Drude spectral density $J(\omega)=\frac{2\Lambda\omega\omega_c}{\omega^2+\omega_c^2}$ with $\Lambda=0.2\,\mathrm{meV}$ and $\omega_c=30\,\mathrm{meV}$.}
\end{figure*}

The Bloch-Redfield equations in the Markovian approximation are,
\begin{equation}\label{eq:redfield}
	\frac{d\rho_{ab}(t)}{dt} = -i\omega_{ab}\rho_{ab}(t) +\sum_{c,d}R_{ab,cd}\rho_{cd}(t),
\end{equation}
where $\rho_{ij}=\bra{i}\hat{\rho}\ket{j}$ are matrix elements of the system's density operator with $\ket{i}$ and $\ket{j}$ being the eigenstates of $\hat{H}_{\mathrm{e}}$ with Bohr frequencies $\omega_{ij}=\omega_i-\omega_j$, and $R_{ab,cd}$ is the Redfield tensor. The Redfield tensor
\begin{equation}
\begin{aligned}
R_{ab,cd}=&-\Big(\delta_{db}\sum_k \Gamma^+_{ak,kc}-\Gamma^{+}_{db,ac}-\Gamma
^{-}_{db,ac}+\delta_{ac}\sum_k \Gamma^-_{dk,kb}\Big)
\end{aligned}
\end{equation}
where
\begin{equation}
\begin{aligned}
\Gamma^+_{ab,cd}=&\frac{1}{\hbar^2}\bra{a}\hat{V}_{\mathrm{e}}\ket{b}\bra{c}\hat{V}_{\mathrm{e}}\ket{d}\tilde{C}(\omega_{dc})\\
\Gamma^-_{ab,cd}=&\frac{1}{\hbar^2}\bra{a}\hat{V}_{\mathrm{e}}\ket{b}\bra{c}\hat{V}_{\mathrm{e}}\ket{d}\tilde{C}(\omega_{ab}),
\end{aligned}
\end{equation}
and $\ket{a}, \ket{b}, \ket{c}$, and $\ket{d}$ are eigenstates of $\hat{H}_{\mathrm{e}}$. The spectral representation of the bath correlation function
\begin{equation}
\begin{aligned}
\tilde{C}(\omega)=&\int_0^{\infty}d\tau e^{i\omega \tau}\Tr[\hat{B}(\tau)\hat{B}(0)\hat{\rho}_b],\\
\Re[\tilde{C}(\omega)]=&\pi \hbar J(|\omega|)\Big[n(|\omega|)+\Theta(\omega)\Big],\\
\Im[\tilde{C}(\omega)]=&\mathcal{P}\int_{0}^{\infty}d\omega' J(\omega') \Big[\frac{n(\omega')+1}{\omega-\omega'}+\frac{n(\omega')}{\omega+\omega'}\Big].
\end{aligned}
\end{equation}
The $\Im[\tilde{C}(\omega)]$ term in the Redfield tensor results in Lamb shifts in the bare Hamiltonian and does not directly affect relaxation, so we ignore it. We define $S_{\mathrm{JT}}(\omega)=\Re[\tilde{C}(\omega)]=\pi \hbar J(|\omega|)\Big[n(|\omega|)+\Theta(\omega)\Big]$ and use a Drude-Lorentz spectral density $J(\omega)=\frac{2\Lambda\omega\omega_c}{\omega^2+\omega_c^2}$ with $\Lambda=0.2\,\mathrm{meV}$ and $\omega_c=30\,\mathrm{meV}$; $S_{\mathrm{JT}}(\omega)$ for this set of parameters is plotted in Fig.~\ref{fig:figS1}. We have $S_{\mathrm{JT}}(0)=2\pi\Lambda k_B T/\omega_c$. For numerical results with the Bloch-Redfield equation plotted in Fig. 3 of the main text and Fig. S2, the equations of motion in Eq. \ref{eq:redfield} were propagated using the adaptive BDF integrator (using solve\_ivp in SciPy) with relative and absolute tolerances of $10^{-8}$ and $10^{-10}$, respectively. We do not make the secular approximation for the numerical calculations.

Representing $\hat{V}_{\mathrm{e}}$
in the eigenbasis $\ket{^1+_{+}}$, $\ket{^{3\gg 1}\mathrm{CT}_{+}}$, $\ket{^1-_{+}}$, $\ket{^1+_{-}}$, $\ket{^{3\gg 1}\mathrm{CT}_{-}}$, and $\ket{^1-_{-}}$ of $\hat{H}_{\mathrm{e}}$, where the states are labeled by the numbers $1$ through $6$, we have
\begin{equation}
    \begin{aligned}
        \hat{V}_{\mathrm{e}} =
\begin{pmatrix}
0 & 0 & 0  & V_{14} & V_{15} & V_{16} \\
0 & 0 & 0  & V_{24} & V_{25} & V_{26} \\
0 & 0 & 0  & V_{34} & V_{35} & V_{36} \\
V_{41} & V_{42} & V_{43} & 0 & 0 & 0 \\
V_{51} & V_{52} & V_{53} & 0 & 0 & 0 \\
V_{61} & V_{62} & V_{63} & 0 & 0 & 0 
\end{pmatrix}
    \end{aligned}
\end{equation}
with
\begin{equation}
    \begin{aligned}
        V_{14}=&\frac{1}{2}F+\frac{\Delta/2}{2\Omega}G \\
        V_{15}=&f_{\mathrm{CT}}\frac{v_0v_{\mathrm{SOC}_+}}{v_0^2+v_{\mathrm{SOC}_+}^2}\sqrt{2+\frac{\Delta}{\Omega}}\\
        V_{16}=&-\frac{\sqrt{v_0^2+v_{\mathrm{SOC}_+}^2}}{2\Omega}G\\
        V_{24}=&-V_{15}\\
        V_{25}=&f_{\mathrm{CT}}\frac{v_0^2-v_{\mathrm{SOC}_+}^2}{v_0^2+v_{\mathrm{SOC}_+}^2}\\
        V_{26}=&f_{\mathrm{CT}}\frac{v_0v_{\mathrm{SOC}_+}}{v_0^2+v_{\mathrm{SOC}_+}^2}\sqrt{2-\frac{\Delta}{\Omega}}\\
        V_{34}=&V_{16}\\
        V_{35}=&-V_{26}\\
        V_{36}=&\frac{1}{2}F-\frac{\Delta/2}{2\Omega}G\\
    \end{aligned}
\end{equation}
where
\begin{subequations}
    \begin{equation}
F=f_{\mathrm{D}}+f_{\mathrm{CT}}\frac{v_0^2-v_{\mathrm{SOC}_+}^2}{v_0^2+v_{\mathrm{SOC}_+}^2},
    \end{equation}
    \begin{equation}
    G=-f_{\mathrm{D}}+f_{\mathrm{CT}}\frac{v_0^2-v_{\mathrm{SOC}_+}^2}{v_0^2+v_{\mathrm{SOC}_+}^2}.
    \end{equation}
\end{subequations}

\begin{figure*}
	\includegraphics[width=\linewidth]{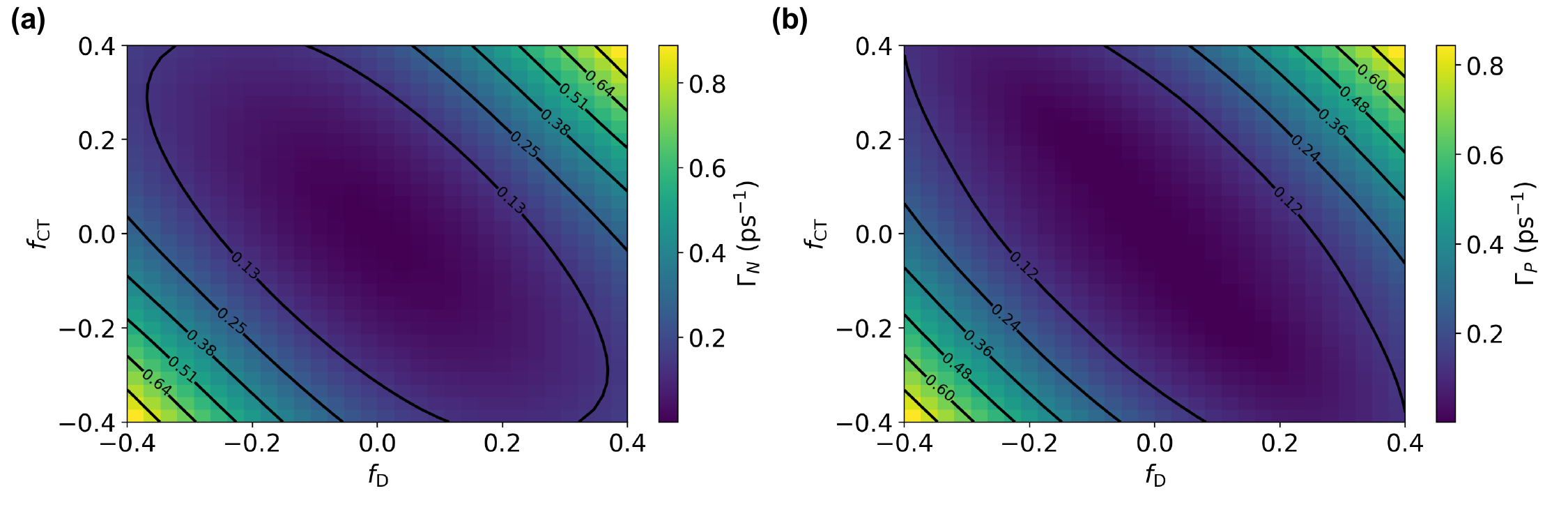}
	\caption{\label{fig:figS2} Relaxation rates. The decay rate (a) of the population difference between transient enantiomers, $\Gamma_{N}$, and (b) the spin-polarization, $\Gamma_{P}$, for different values of system-bath coupling parameters $f_{\mathrm{D}}$ and $f_{\mathrm{CT}}$ when the initial state is the singlet donor $\ket{^1\mathrm{D}_{+}}$.}
\end{figure*}

Our initial density matrix $\hat{\rho}(0)=\ket{^1\mathrm{D}_{+}}\bra{^1\mathrm{D}_{+}}$ in the $\hat{H}_{\mathrm{e}}$ eigenbasis would only contain non-zero density matrix elements $\rho_{11}$, $\rho_{33}$, $\rho_{13}$, and $\rho_{31}$. When $|v_0|\gg |v_{\mathrm{SOC}_+}|$, the terms $V_{15}$, $V_{35}$, $V_{24}$, and $V_{26}$ will be much smaller than others; as a result, starting with populations $\rho_{11}$ and $\rho_{33}$ will lead to populations $\rho_{44}$ and $\rho_{66}$, but far lower population would flow into $\rho_{55}$ and $\rho_{22}$. For our parameters, $v_0=75$meV, $v_{\mathrm{SOC}_{\pm}}=\pm25$meV, and $\Delta=-30$meV, we have $\Omega\approx80$meV. Therefore, the energy differences between eigenstates of $\hat{H}_{\mathrm{e}}$ are $0$meV, $\sim 80$meV and $\sim 160$meV. From Fig.~\ref{fig:figS1}, we see that $S_{\mathrm{JT}}(\omega)$ is about four times larger for $\omega=0$meV than $\omega=80$meV and $\omega=160$meV, so we can focus our attention on terms of the Redfield tensor that involve $V_{14}$, $V_{25}$ and $V_{36}$ as they involve degenerate states. When both $f_{\mathrm{D}}$ and $f_{\mathrm{CT}}$ are positive or negative, the $F$ dominates both $V_{14}$ and $V_{36}$ and we expect the relaxation rate to be $\propto F^2$ as can be clearly seen from the contour lines in Fig.~\ref{fig:figS2} and from Fig. 3(c) in the main text. When $f_{\mathrm{D}}\approx -f_{\mathrm{CT}}\frac{v_0^2-v_{\mathrm{SOC}_+}^2}{v_0^2+v_{\mathrm{SOC}_+}^2}$, we expect the relaxation rate to be $\propto G^2$ and slower as $F\approx0$ and $G$ has an additional coefficient of $|\Delta /2\Omega|$ in $V_{14}$ and $V_{36}$. We can qualitatively observe this from the dark blue region in Fig.~\ref{fig:figS2}(a-b); however, we have not quantified this, as we need to run longer simulations to obtain reliable $\Gamma_{N}$ and $\Gamma_{P}$ in this region because the relaxation is slow.




The population difference $\langle\Delta \hat{N}\rangle=\rho_{11}+\rho_{22}+\rho_{33}-\rho_{44}-\rho_{55}-\rho_{66}$ in the eigenbasis of $\hat{H}_{\mathrm{e}}$. As $S_{\mathrm{JT}}(\omega)$ is larger when $\omega=0$ than at other Bohr frequencies of $\hat{H}_{\mathrm{e}}$ for our parameters, within the partial-secular approximation, we only need to consider population transfer between degenerate $\lambda=\pm$ eigenstates with $\omega_{1}=\omega_4$, $\omega_2=\omega_5$, and $\omega_3=\omega_6$. This would give us an approximate equation of motion for
 $\Delta\hat{N}$,
\begin{equation}\label{eq:dNeq}
\begin{aligned}
       \frac{d\langle\Delta \hat{N}\rangle}{dt}\approx&-2R_{11,44}(\rho_{11}-\rho_{44})-2R_{22,55}(\rho_{22}-\rho_{55})-2R_{33,66}(\rho_{33}-\rho_{66})\\
       \approx&-\frac{S_{JT}(0)}{\hbar^2}F^2\langle\Delta\hat{N}\rangle\\
       =&-\frac{2\pi k_B T}{\hbar}\frac{\Lambda}{\hbar\omega_c}F^2\langle\Delta\hat{N}\rangle,
\end{aligned}
\end{equation}
 where have ignored $R_{22,55}(\rho_{22}-\rho_{55})$ as initially there is no population in the states involved $\rho_{22}(0)=\rho_{55}(0)=0$ and we have also assumed $|\Delta G/2 \Omega|\ll |F|$, so $V_{14}\approx V_{36}\approx F/2$.